\documentclass[12pt,twocolumn,tighten]{aastex62}
\usepackage{amsmath,amstext,amssymb}
\usepackage[T1]{fontenc}
\usepackage{apjfonts}
\usepackage[figure,figure*]{hypcap}
\usepackage{graphics,graphicx}
\usepackage{hyperref}
\usepackage{comment}

\received{August 18, 2019}
\revised{September 20, 2019}
\accepted{September 23, 2019}
\submitjournal{{\it The Astrophysical Journal Supplement}}

\shortauthors{Bouma et al.}
\shorttitle{CDIPS I}

\newcommand{\sVInumberlcs}{67{,}612\ }  
\newcommand{\sVIInumberlcs}{91{,}731\ }  
\newcommand{\numberlcs}{159{,}343\ } 
\newcommand{\numberclusters}{596\ } 

\newcommand{\stscilink}{\textsc{\url{archive.stsci.edu/hlsp/cdips}}}

\newcommand{\datasetlink}{\textsc{\dataset[doi.org/10.17909/t9-ayd0-k727]{https://doi.org/10.17909/t9-ayd0-k727}}}

\listofchanges

\begin{document}

\title{
  Cluster Difference Imaging Photometric Survey. I.
  Light Curves of Stars in Open Clusters from TESS Sectors 6 \& 7
}

\correspondingauthor{L. G. Bouma}
\email{luke@astro.princeton.edu}

\author[0000-0002-0514-5538]{L. G. Bouma}
\affiliation{ Department of Astrophysical Sciences, Princeton
University, 4 Ivy Lane, Princeton, NJ 08540, USA}
\author[0000-0001-8732-6166]{J. D. Hartman}
\affiliation{ Department of Astrophysical Sciences, Princeton
University, 4 Ivy Lane, Princeton, NJ 08540, USA}
\author[0000-0002-0628-0088]{W. Bhatti}
\affiliation{ Department of Astrophysical Sciences, Princeton
    University, 4 Ivy Lane, Princeton, NJ 08540, USA}
\author[0000-0002-4265-047X]{J. N. Winn}
\affiliation{ Department of Astrophysical Sciences, Princeton
University, 4 Ivy Lane, Princeton, NJ 08540, USA}
\author[0000-0001-7204-6727]{G. \'A. Bakos}
\affiliation{ Department of Astrophysical Sciences, Princeton
University, 4 Ivy Lane, Princeton, NJ 08540, USA}

\begin{abstract}
  The Transiting Exoplanet Survey Satellite (TESS) is providing
  precise time-series photometry for most star clusters in the solar
  neighborhood.
  Using the TESS images, we have begun a Cluster Difference Imaging
  Photometric Survey (CDIPS), in which we are focusing both on stars
  that are candidate cluster members, and on stars that show
  indications of youth.
  Our aims are to discover giant transiting planets with known ages,
  and to provide light curves suitable for studies in stellar
  astrophysics.
  For this work, we made \numberlcs light curves of candidate young
  stars, across \numberclusters distinct clusters.  Each light curve
  represents between 20 and 25 days of observations of a star brighter
  than $G_{\rm Rp}=16$, with 30-minute sampling.
  We describe the image subtraction and time-series analysis
  techniques we used to create the light curves, which have noise
  properties that agree with theoretical expectations.
  We also comment on the possible utility of the light curve sample
  for studies of stellar rotation evolution, and binary eccentricity
  damping.
  The light curves\footnote{\stscilink}, which cover about one
  sixth of the galactic plane, are available as a High Level Science
  Product at MAST: \datasetlink.
\end{abstract}



\section{Introduction}
\label{sec:intro}

Each of the several thousand star clusters of the Milky Way is a gift
to astrophysics, providing a sample of stars that vary widely in mass
but all have approximately the same age and composition.  Time-series
photometry of clusters has many applications. By measuring rotation
periods over a range of ages, we can study the angular momentum
evolution of stars and improve our ability to determine stellar ages
through gyrochronology \citep[{\it
e.g.},][]{skumanich_time_1972,barnes_color-period_2015,meibom_spin-down_2015,curtis_tess_2019}.
By measuring the eccentricity distribution of binary stars as a
function of age, we can study the tidal circularization process
\citep{meibom_robust_2005,milliman_wiyn_2014,price-whelan_binary_2018}.
The detection of eclipsing binaries (EBs) can also lead to the precise
determination of the absolute dimensions of the stars and stringent
tests of stellar-evolutionary models
\citep{luhman_formation_2012,stassun_review_2014,kraus_mass-radius_2015}.
Finally, transiting exoplanets discovered in clusters can shed light
on the timescales for processes in planet formation, evolution, and
migration
\citep[][]{Fortney_et_al_2007,Mann_K2_33b_2016,David_et_al_2017}, as
well as on the effects of metallicity
\citep[][]{fischer_planet-metallicity_2005,petigura_metallicity_2018}.

The Transiting Exoplanet Survey Satellite (TESS,
\citealt{ricker_transiting_2015}) holds the promise to deliver the
most homogeneous and comprehensive cluster photometric survey in
history.  Based on the cluster membership data of
\citet{Kharchenko_et_al_2013}, approximately $2\times10^5$ open
cluster members brighter than $T=16$ will be observed in the
full-frame images (FFIs) over the first two years of TESS.  This count includes the
\citeauthor{Kharchenko_et_al_2013} ``most probable'' members, which
are stars
with kinematic, photometric, and spatial membership
probabilities each independently exceeding 61\% \citep{kharchenko_global_2012}.  The
actual number of stars in clusters is likely larger, as the membership
catalogs are not yet complete, even within the nearest kiloparsec
\citep[{\it
e.g.},][]{roser_nine_RSG_2016,cantat-gaudin_gaia_2018,cantat-gaudin_newOCs_2019,kounkel_untangling_2019,sim_open_2019}.

A major barrier to deriving precise photometry from the TESS images is
the relatively poor angular resolution ($21''$ per pixel).   The
problems with crowding and complex backgrounds are so severe that the
TESS Candidate Target List deprioritizes 2-minute targets within
$15^\circ$ of the galactic plane\footnote{During the first year of
TESS observations, objects within $15^\circ$ of the galactic plane
were deprioritized. Starting in the second year of observations, the
cutoff changed to objects within $10^\circ$ of the galactic plane
\citep{stassun_TIC8_2019}.}~\citep{stassun_TIC_2018,stassun_TIC8_2019}.
This decision was made because the large pixel size and the high
stellar surface density make aperture photometry unreliable. 
By consequence, most stars in clusters, which are usually near
the galactic plane, will go unprocessed by the official {\it TESS}
data reduction pipeline.

One way to quantify the blending problem is to determine what fraction
of the total flux in a photometric aperture is contributed by a
particular target star. Aperture photometry is reliable when this
fraction is close to unity.  Difference imaging
\citep{Alard_Lupton_1998,miller_optimal_2008}, in our group's
experience, can be viable down to crowding fractions of 10\%.  Based
on the \citet{Kharchenko_et_al_2013} cluster membership data and the
density of background stars, we calculated that the median dilution
fraction is 0.13, for cluster stars with $T< 16$ and an aperture
radius of 2 pixels.  Thus, for at least $\sim$$10^5$ cluster stars,
difference imaging may be advantageous.

Difference imaging avoids the primary effects of blending through
forced-aperture photometry.  In this method, the pixel coordinates of
stars are calculated from an astrometric solution, and the reference
fluxes are determined from a calibrated catalog magnitude-to-flux
relation (see \S~\ref{subsubsec:photref}).  The deviation from the
reference flux is measured on the difference image.  Assuming that
only a single source is variable, blended neighbor stars only act to
increase Poisson noise, down to the angular resolution of the source
catalog used to determine the reference flux.  This is a major benefit
of performing image subtraction in crowded fields.

We have therefore begun to apply difference imaging to the TESS
FFIs, with a focus on any star that could be a cluster member.
We are also including some stars that we suspect are young due to
combined photometric and astrometric indicators.  A major motivation
for this effort is to discover giant transiting planets with known
ages.  The focus of the present study however is to describe our
methods, and to produce a general-purpose dataset applicable for
studies both in exoplanetary and also stellar astrophysics.

For the remainder of this work, we adopt the term ``star cluster'', or
simply cluster, to refer to a coeval group of stars.  This includes
open clusters, as well as the moving groups and stellar associations
that have been discovered since the late 1990s
\citep{zuckerman_young_2004}. 

The plan of action is as follows. In \S~\ref{sec:starselection}, we
describe how target stars were selected. We then present the
photometric and image processing methods we used to produce light
curves for these stars (\S~\ref{sec:method}).  In TESS Sectors 6 and
7, these methods yielded \numberlcs light curves, and in
\S~\ref{subsec:lcstatistics} we summarize their statistical
properties.  We then give an example of how to use the data to identify
pulsating stars, eclipsing binaries, and transiting planets
(\S~\ref{subsec:identifying_variability}).  We close in
\S~\ref{sec:conclusion} with a summary of our findings, and discuss them
with an eye towards the studies we hope this and further data
processing will enable.

\section{Method: Star Selection}
\label{sec:starselection}

\begin{figure*}[!t]
	\begin{center}
		\leavevmode
		\includegraphics[width=0.9\textwidth]{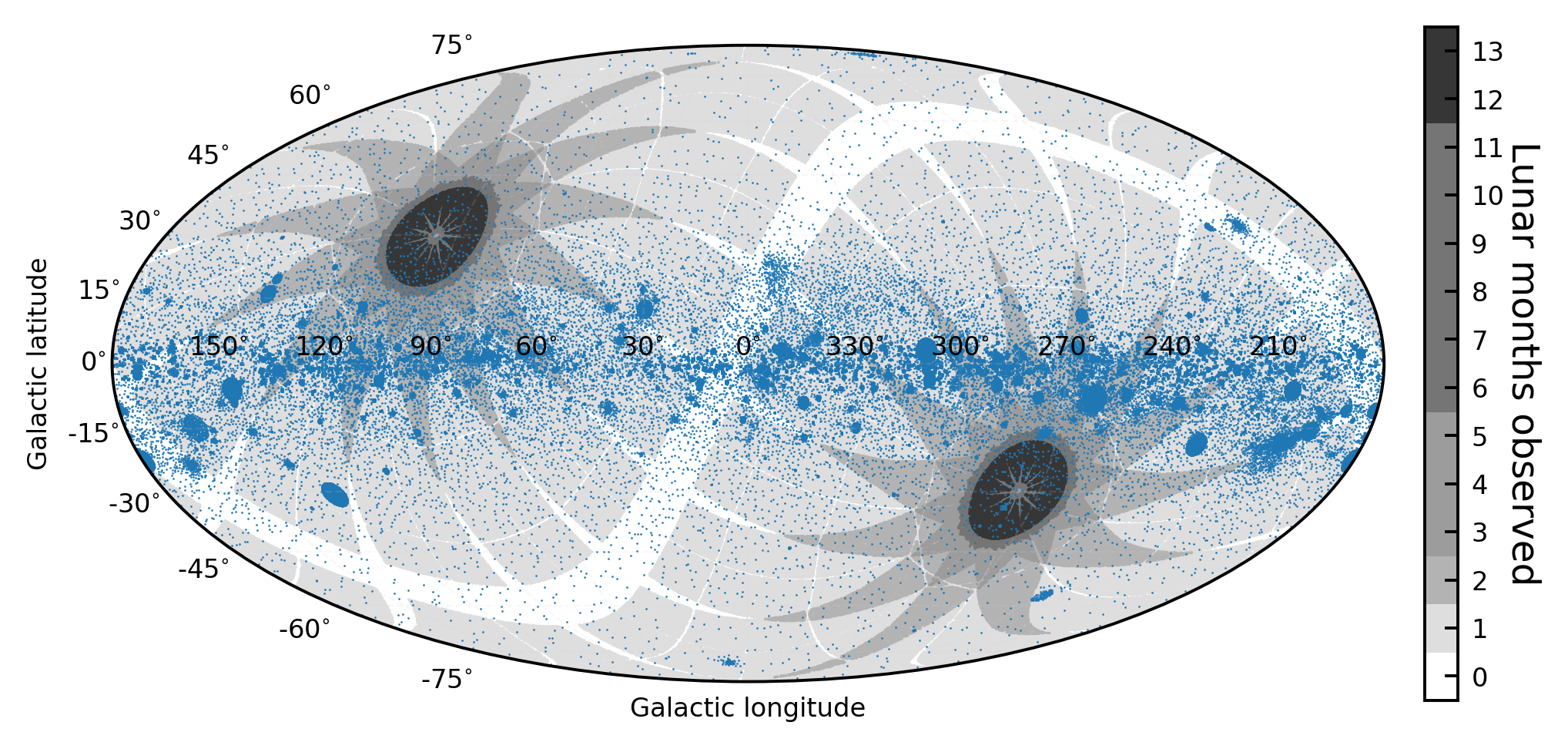}
	\end{center}
	\vspace{-0.5cm}
	\caption{
    Target star positions (blue) and nominal TESS observing footprint
    (gray).  Target stars are either candidate members of clusters, or
    else have other youth indicators (see \S~\ref{sec:starselection}).
    Most will be observed for one or two lunar months during the TESS
    Prime Mission.  Camera 1 is centered at $(l,b)=(203^\circ,
    -6^\circ)$ and $(218^\circ,15^\circ)$ in Sectors 6 and 7,
    respectively.
    \label{fig:cdips_targets_positions}
	}
\end{figure*}

\begin{figure}[!t]
	\begin{center}
		\leavevmode
		\gridline{\fig{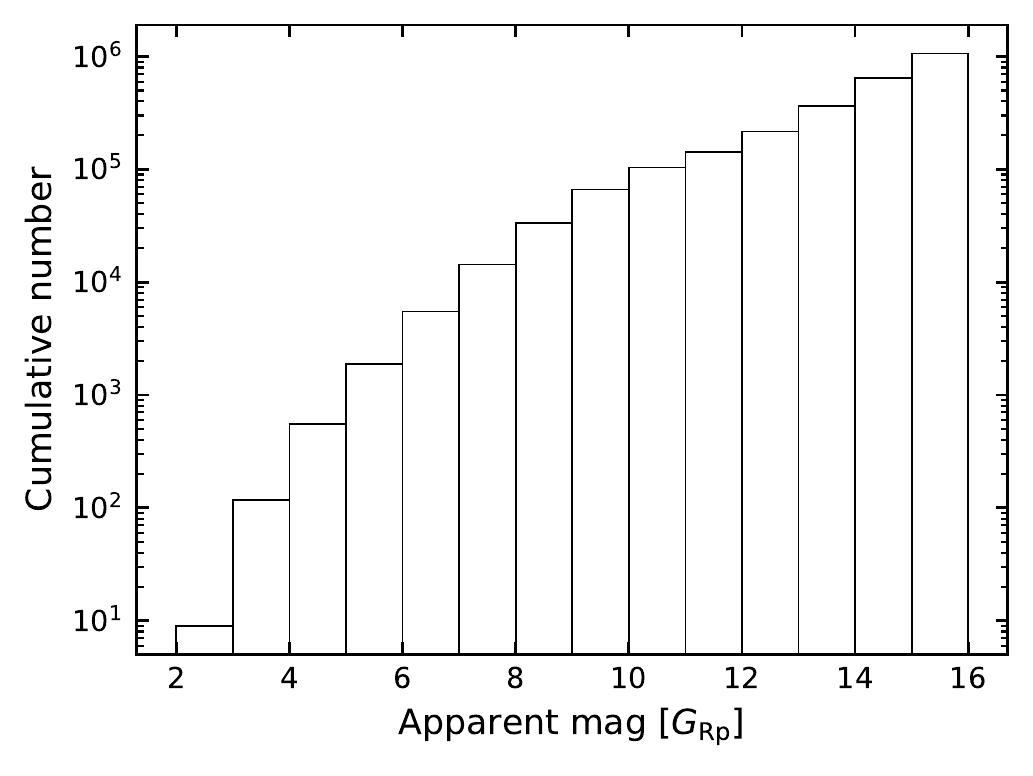}{0.42\textwidth}{}}
		\vspace{-0.8cm}
		\gridline{\fig{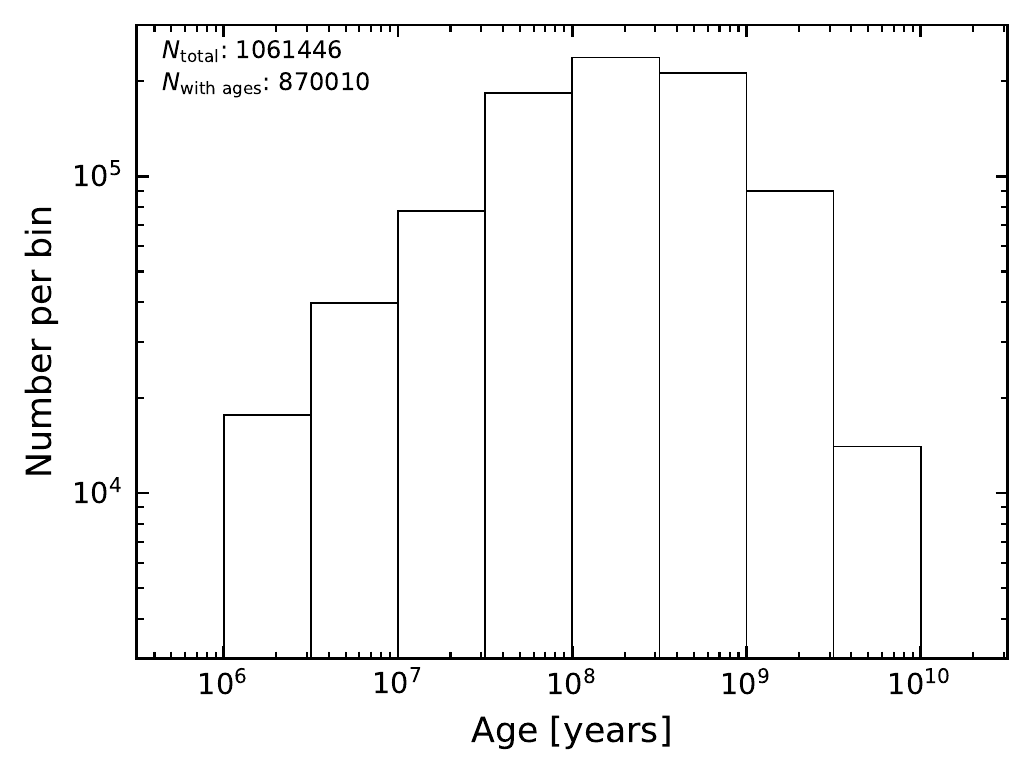}{0.42\textwidth}{}}
		\vspace{-0.8cm}
		\gridline{\fig{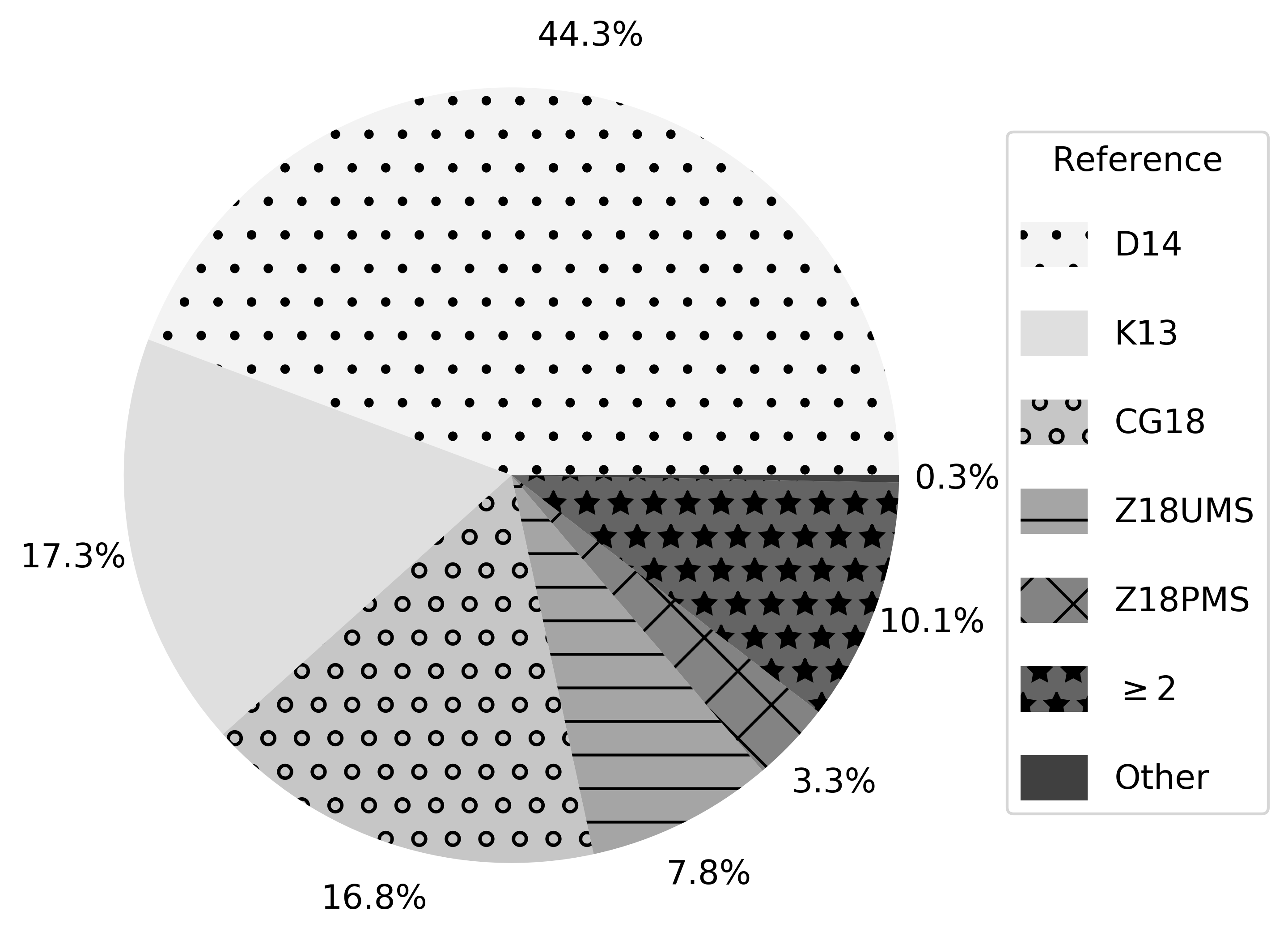}{0.42\textwidth}{}}
	\end{center}
	\vspace{-0.8cm}
	\caption{
		Target star statistics.
		{\it Top.} Cumulative counts as a function of apparent Gaia $Rp$-band
		magnitude.  
		{\it Middle.} Histogram of target star ages, for the subset of
		stars with ages matched against \citet{Kharchenko_et_al_2013}.
		{\it Bottom.} Provenance of cluster membership.  Percentages are
		relative to the $N_{\rm total}=1{,}061{,}446$ target stars, which are listed in
		Table~\ref{tbl:cdips_targets}. Symbols
		are as follows.
		D14 is \citet{dias_proper_2014}.
		K13 is \citet{Kharchenko_et_al_2013}.
		CG18 is \citet{cantat-gaudin_gaia_2018}.
		Z18 is \citet{zari_3d_2018}, with upper main-sequence and
		pre-main-sequence samples sub-divided.
		``$\geq 2$'' indicates at least two authors reported a star as a
		candidate cluster member.
		\label{fig:cdips_targets}
	}
\end{figure}

\begin{figure*}[!t]
	\gridline{\fig{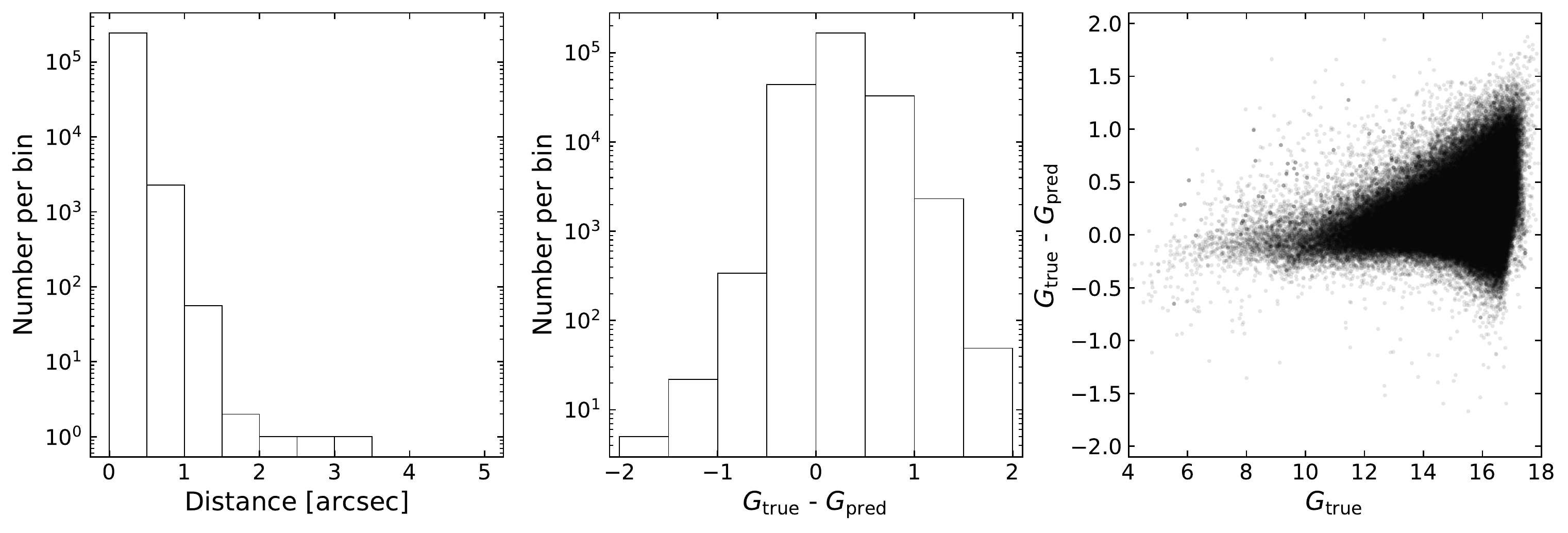}{0.83\textwidth}{}}
	\vspace{-1.1cm}
	\gridline{\fig{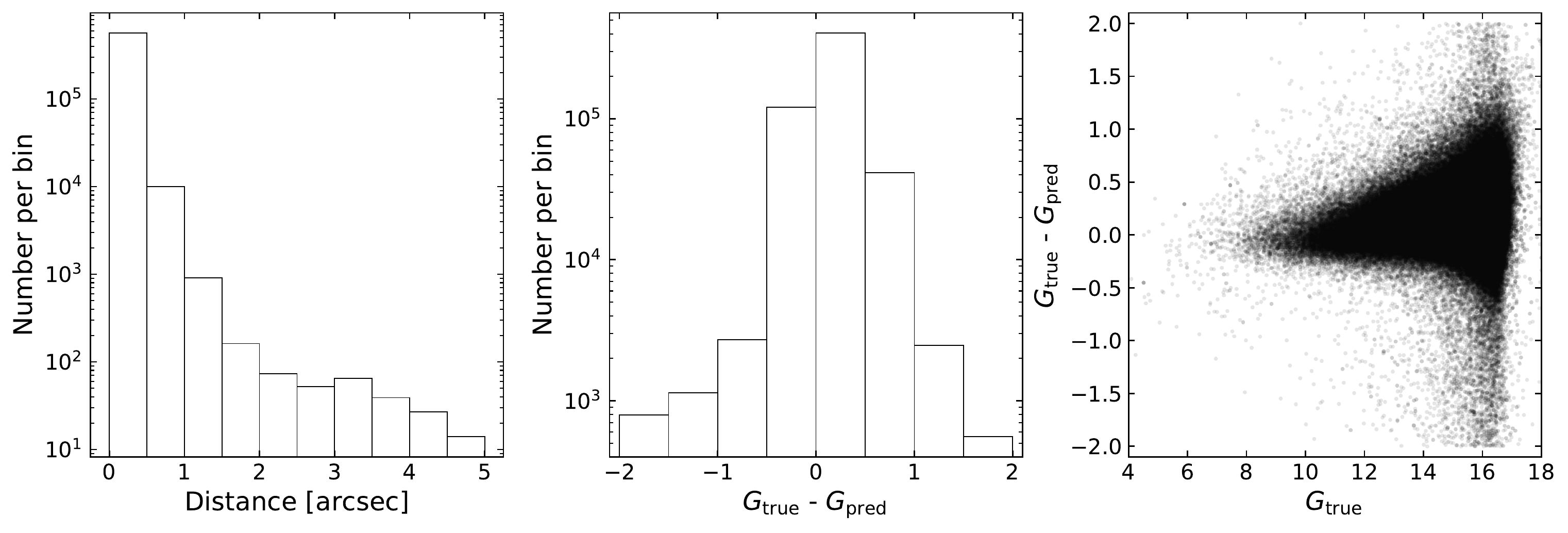}{0.83\textwidth}{}}
	\vspace{-0.8cm}
	\caption{
    {\it Top.} Quality diagnostics from cross-matching
    \cite{Kharchenko_et_al_2013} cluster members against Gaia-DR2.  A
    histogram of the distances between matched stars is on the left; a
    histogram of the difference between the true $G$-band magnitude
    and that predicted from 2MASS photometry is in the middle; a
    scatter plot of the same magnitude difference as a function of
    $G$-band magnitude is on the right.
		{\it Bottom.} Same, but cross-matching \cite{dias_proper_2014}
		cluster members to Gaia-DR2.
	}
	\label{fig:xmatch_info}
\end{figure*}

The main aim of the CDIPS project is to increase the number of cluster
stars for which photometric time-series are available, and thereby
facilitate studies of exoplanetary and stellar processes across
different times and stellar environments.  A key step is therefore to
define a sample of stars that are thought to be young, or members of
clusters, or both.

A homogeneous membership calculation for every known cluster is a
large undertaking, and currently falls outside our scope.  So too is a
homogeneous search for young stars across the galaxy.  Instead, we
have opted to collect and concatenate catalogs from the literature.
We then use the resulting meta-catalog to identify target stars within
the TESS images.

The criteria for inclusion in our target star list are necessarily
heterogeneous across different catalogs.  We aim for completeness, not
accuracy.  If there has been a claim in the literature that a star
should be considered a cluster member, or a young star, we would like
to err on the side of reporting a light curve for the star.  For stars
that are photometrically interesting, we can then perform post-hoc
quality checks using Gaia-DR2 astrometry and photometry to assess
cluster membership and youth.

First, we describe the catalogs we used to identify members of open
clusters (\S~\ref{subsec:oc}).  Then, we discuss the catalogs we used
to identify members of moving groups, stellar associations, and more
generally young stellar populations (\S~\ref{subsec:mg}).  In
\S~\ref{subsec:ocmgsummary} we give
summary statistics for the entire sample of about one million target
stars, and we list the targets in Table~\ref{tbl:cdips_targets}.

\subsection{Big catalogs: open clusters}
\label{subsec:oc}

At the time of our analysis, two relatively large, homogeneous cluster
memberships studies had been performed using Gaia-DR2: those by
\citet{cantat-gaudin_gaia_2018} and \citet{gaia_hr_2018}.  There were
also two large membership studies pre-dating Gaia-DR2 based on proper
motion and photometric catalogs: the studies of
\citet{Kharchenko_et_al_2013} and \citet{dias_proper_2014}.

\paragraph{{\it Gaia}-derived OC memberships}

\citet{cantat-gaudin_gaia_2018} used an unsupervised membership
assignment algorithm \citep{krone-martins_upmask_2014} to identify
cluster members using Gaia-DR2 positions, proper motions, and
parallaxes.  They used Gaia photometry and radial velocities to verify
the membership claims.  From their Table~2, we collected 401{,}448
cluster members, in 1229 clusters, down to their limiting magnitude of
$G=18$.

\citet{gaia_hr_2018} reported members of a smaller, more select group
of well-studied open clusters. From their Table~A1, we collected
40{,}903 cluster members, in 41 open clusters, mostly within
$500\,{\rm pc}$. While this work also included memberships for
globular clusters, we omitted them from consideration.

Given the high quality of Gaia-DR2 astrometry, these two membership
sources are our most reliable sources of membership information.  In
our photometric reduction,  our default identifier for all sources is
correspondingly the Gaia-DR2 \texttt{source\_id}.  The TESS Input
Catalog (TIC) data for each target star are then collected using the
Gaia-DR2 source identifier \citep{stassun_TIC_2018,stassun_TIC8_2019}.  

\paragraph{Pre-{\it Gaia} OC memberships}

\citet{Kharchenko_et_al_2013} used proper motions calculated in PPMXL
\citep[][a combination of USNO-B1{.}0 and 2MASS
astrometry]{roeser_ppmxl_2010} and near-infrared photometry from 2MASS
\citep{skrutskie_tmass_2006} to report the existence of 2859 open
clusters and stellar associations. We omitted globular clusters by
excluding any entry of type `\texttt{g}'.  We selected the most
probable cluster members (``$1\sigma$ members'') using the combined
photometric, kinematic, and spatial criteria described by
\citet{kharchenko_global_2012}.  Then, to obtain {\it
Gaia}-DR2 source identifiers for the members, we performed a
crossmatch for Gaia-DR2 sources within 5 arcseconds of the listed
positions.  To improve the quality of the cross-match, we used the
2MASS photometry to predict the $G$-band magnitudes\footnote{See
\url{https://gea.esac.esa.int/archive/documentation/GDR2/Data_processing/chap_cu5pho/sec_cu5pho_calibr/ssec_cu5pho_PhotTransf.html},
(accessed \texttt{2019-03-29}), or \citet{carrasco_gaia_2016}}, and
required that the measured $G$-magnitude fall within 2 magnitudes of
the predicted $G$-magnitude.  If multiple neighbors matched the
position and magnitude constraints, we took the nearest spatial
neighbor as the match.  From 373{,}226 stars, this yielded a unique
best neighbor for 352{,}332 stars (94.4\% of the sample), and a choice
between two neighbors for 17{,}774 stars. 

The second (non-{\it Gaia} derived) open cluster membership catalog we
used was the \citet{dias_proper_2014} catalog, which was based on
UCAC4 proper motions acquired by the US Naval Observatory
\citep{zacharias_fourth_2013}.  From their 1805 reported open
clusters, we selected sources with quoted membership probability above
50\%.  To obtain Gaia-DR2 source identifiers for the members, we
performed a similar crossmatch, looking for sources within 5
arcseconds of the listed positions, and within $\pm$2 $G$-band
magnitudes of the prediction.  From 2{,}034{,}269 stars, this yielded
a unique best neighbor for 1{,}828{,}630 stars (89.9\% of the sample),
and a choice between two neighbors for 8.7\% of the remaining sample. 

The distributions of various cross-matching statistics are shown in
Figure~\ref{fig:xmatch_info}.  The distances between matches is
typically below 1 arcsecond.  The \citeauthor{dias_proper_2014}
catalog shows stronger crowding effects at the faint end than
the \citeauthor{Kharchenko_et_al_2013} catalog, and likely has a
larger number of false matches.  At their faint ends, both catalogs
show a tendency for true $G$-band magnitudes to be larger than
predicted $G$-band magnitudes, presumably due to dust reddening.

\subsection{Smaller catalogs: moving groups and stellar associations}
\label{subsec:mg}

Stars in moving groups and stellar associations are interesting for
similar reasons as stars in open clusters.  
Relative to open cluster members though, stars in moving groups are 
usually closer, brighter, and
exist in less crowded environments.

To identify stars in these types of groups, we matched the following studies
against Gaia-DR2:
\citet{gagne_banyan_XII_2018},
\citet{gagne_banyan_XI_2018},
\citet{gagne_banyan_XIII_2018},
\citet{kraus_tucanahor_2014},
\citet{roser_deep_2011}, 
\citet{bell_32ori_2017},
\citet{rizzuto_multidimensional_2011},
\citet{oh_comoving_2017}, and
\citet{zari_3d_2018}. The methods applied in these studies
vary from kinematic analyses, to astrometric analyses included
Gaia-DR1 parallaxes, to photometric searches for infrared excesses, to
spectroscopic studies including RVs, H$\alpha$
emission, and Li absorption.

For the Gagn\'e et al{.}~catalogs, a large number of the stars have
high proper motions.  However, some of the stars do not have reported
proper motions.  To perform the cross-match, we searched the Gaia-DR2
archive for sources within 10 arcseconds of the listed positions
(propagated to the Gaia-DR2 J2015.5 epoch, if the proper motions were
available, otherwise simply using the listed J2000 positions).  We
also imposed a $G<18$ cut on any putative matches.  We then chose the
nearest neighbor by spatial separation.  Of 3012 moving group members
collected from the three combined Gagn\'e et al{.} catalogs, this
procedure yielded 2702 matches.

The \citet{kraus_tucanahor_2014}, \citet{roser_deep_2011}, and
\citet{bell_32ori_2017} studies reported members in Tucana-Horologium,
the Hyades, and 32$\,$Ori respectively.  Applying the same procedure
as for the Gagn\'e catalogs gave 187, 684, and 119 matches
respectively, compared to 205, 724, and 141 initially reported
members.  Note that \citet{kraus_tucanahor_2014} found that only
$\sim$70\% of their listed members have spectroscopic indicators
consistent with membership in Tucana-Horologium.

\citet{rizzuto_multidimensional_2011} focused on a single group: the
Sco OB2 association. We used their reported Hipparcos identifiers, and
matched against the {\it Gaia} archive's
\texttt{hipparcos2\_best\_neighbour} table, which gave 319
nearest-neighbor stars from 436 candidate members.

\citet{oh_comoving_2017} searched for comoving stars in the $\approx$2
million stars that appear in both the Tycho-2 and Gaia-DR1
catalogs.  They found many wide binaries, and also identified a large
number of comoving groups.  We chose the 2{,}134 stars that they
reported were in groups with sizes of at least 3 stars.  Using their
Gaia-DR1 source identifiers, we matched against the {\it Gaia}
archive's \texttt{dr1\_neighbourhood} table, which gave 1{,}881
nearest-neigbor stars in groups of at least three stars
\citep{marrese_gaia_2019}.

Finally, \citet{zari_3d_2018} constructed a sample of young stars
within $500\,{\rm pc}$ using data from Gaia-DR2. Two subsamples were
made: an upper main-sequence (MS) sample, with 86{,}102 stars, and
a pre-MS sample, with 43{,}719 stars.  Each was created from a
careful combination of distinct astrometric and photometric cuts.
These stars are the youngest, closest stars, spread across
star-forming complexes in Sco-Cen, Orion, Vela, Taurus, and other
regions of the sky.  Though most of these stars are not directly
identified with moving groups or open clusters, their reported youth
and proximity to star forming regions justifies their inclusion in our
search sample.

\subsection{Summary of target stars}
\label{subsec:ocmgsummary}

After collecting the aforementioned lists, we merged them into a
single table. We queried Gaia-DR2's \texttt{gaia\_source} table to
retrieve each star's apparent $G$, $G_{\rm Rp}$, and $G_{\rm Bp}$
magnitudes, as well as their astrometric measurements $(\alpha,
\delta, \mu_\alpha, \mu_\delta, \pi)$.  Finally, we required that
$G_{\rm Rp} < 16$, which is roughly the level for which the 1-hour
photometric precision of TESS is predicted to be 1\%
\citep{ricker_transiting_2015}.

The resulting CDIPS target star list, consisting of 1{,}061{,}447
unique stars from 13 distinct catalogs, is given in
Table~\ref{tbl:cdips_targets}.  The cumulative distribution of target
star brightnesses, as well as a histogram of the ages, is shown in
Figure~\ref{fig:cdips_targets}.  Relative to field stars, our target
star sample is young, with a most probable age of 100$\,$Myr.

Figure~\ref{fig:cdips_targets} also shows the relative fraction of
stars from each catalog.  The largest number of stars come from
\citealt{dias_proper_2014} (44.3\%), \citealt{Kharchenko_et_al_2013}
(17.3\%), \citealt{cantat-gaudin_gaia_2018} (16.7\%), and
\citealt{zari_3d_2018} (11.1\%, of which 7.8\% are OBA stars, and
3.3\% are pre-MS stars).  107{,}647 of the stars, or about 10\% of the
collection, have cluster memberships reported by multiple authors.
Since the membership probability calculations often use independent
data and methods, agreement between multiple investigators on a given
star's cluster membership is a helpful indication of it being a
member.

Different catalogs have different standards for deciding which stars
are members.  For the \citet{dias_proper_2014} catalog, their
membership calculation included only spatial and kinematic
information, and we used a relatively low probability threshold when
including their stars (based on the criteria
\citealt{dias_proper_2014} used for their star counts).  The
\citet{Kharchenko_et_al_2013} catalog combined spatial, kinematic, and
photometric information to derive their membership probabilities.  We
also used a more restrictive membership probability cut (again,
following the criteria they used for their star counts), so this
sub-sample is likely less contaminated with field interlopers.
\citet{cantat-gaudin_gaia_2018} used spatial, kinematic, and
astrometric information from Gaia-DR2. Despite the lack of photometric
information, the quality of the Gaia data suggest that the field
contamination rate will be lowest for the
\citet{cantat-gaudin_gaia_2018} sample.

To assign unique cluster names, we adopted the name given by
\citet{Kharchenko_et_al_2013} whenever possible.
Appendix~\ref{appendix:uniquenames} describes  how this was done in
detail.  For moving groups not identified by
\citet{Kharchenko_et_al_2013}, we used the constellation-based naming
convention from \citet{gagne_banyan_XI_2018}.  Otherwise, we used the
name reported by the original catalog claiming membership.  This
process reduced 16,425 name variants down to 3,216 unique cluster
names.  Though we have made every effort to avoid duplicates, a small
number may remain, so we advise inspection of the \texttt{cluster}
column as well as the references given in the \texttt{reference}
column rather than using the \texttt{unique\_cluster\_name} column to
analyze individual objects of interest.  Nonetheless, 87.7\% of the
$\sim$million unique target stars are matched to clusters named by
\citet{Kharchenko_et_al_2013}, and 88.8\% are assigned a cluster name.
The remainder are mostly young stars from \citet{zari_3d_2018}.
Ages and their uncertainties were then assigned using the parameters
reported by \citet{Kharchenko_et_al_2013}.

\section{Method: Photometry}
\label{sec:method}

\begin{figure}[!t]
	\begin{center}
		\leavevmode
		\includegraphics[width=0.35\textwidth]{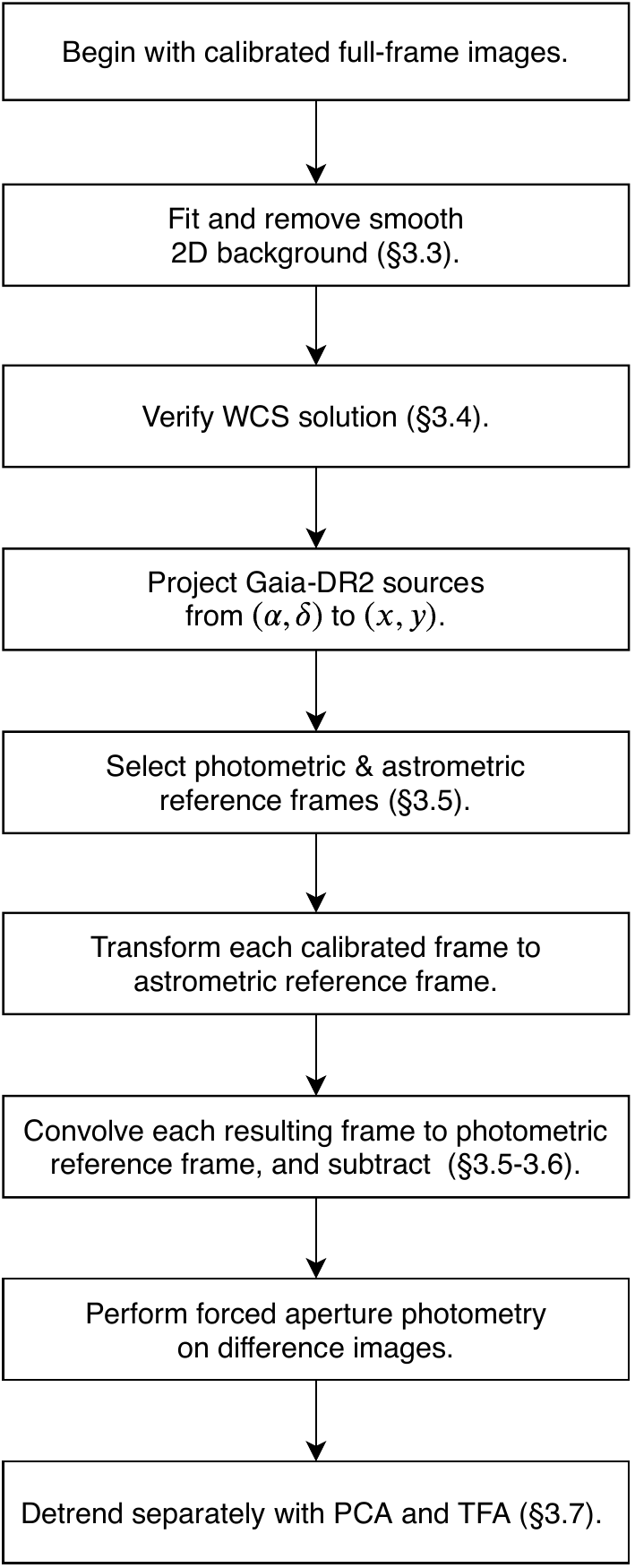}
	\end{center}
	\vspace{-0.2cm}
	\caption{
    Conceptual overview of photometric reduction pipeline.
    Details are given in \S~\ref{sec:method}.
	\label{fig:pipeline}
	}
\end{figure}

\begin{figure*}[!t]
    \begin{center}
        \leavevmode
        \includegraphics[width=0.93\textwidth]{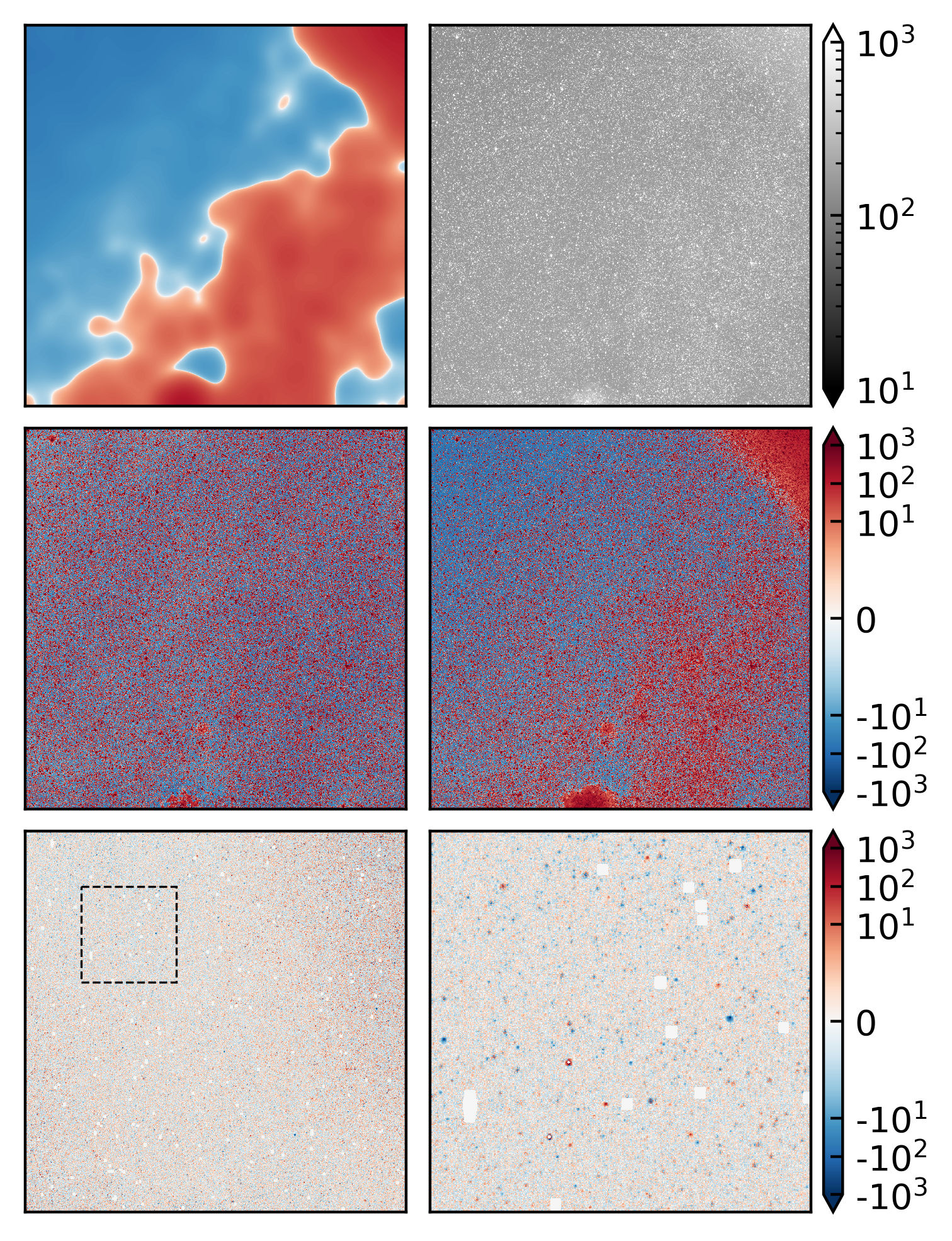}
    \end{center}
    \vspace{-0.9cm}
    \caption{
        Stages of image processing for an image obtained during the middle
        of an orbit. Counts are in ADU.
        {\it Top right}. Calibrated image.
        {\it Top left}. Smooth background estimate.
        {\it Middle right}. Calibrated image minus its median pixel value.
        {\it Middle left}. Calibrated image minus smooth background.
        {\it Bottom left}. Difference image.
        {\it Bottom right}. Zoom of difference image, corresponding to
        the square with dashed lines.
        Each image is $(2048\times2048)$ pixels, except for the
        bottom-right image, which is $(512\times512)$.  All images
        except the top-right have identical color maps.  The sector,
        camera, and CCD are (6, 1, 2).
        \label{fig:stages_good}
    }
\end{figure*}

\begin{figure*}[!t]
    \begin{center}
        \leavevmode
        \includegraphics[width=0.93\textwidth]{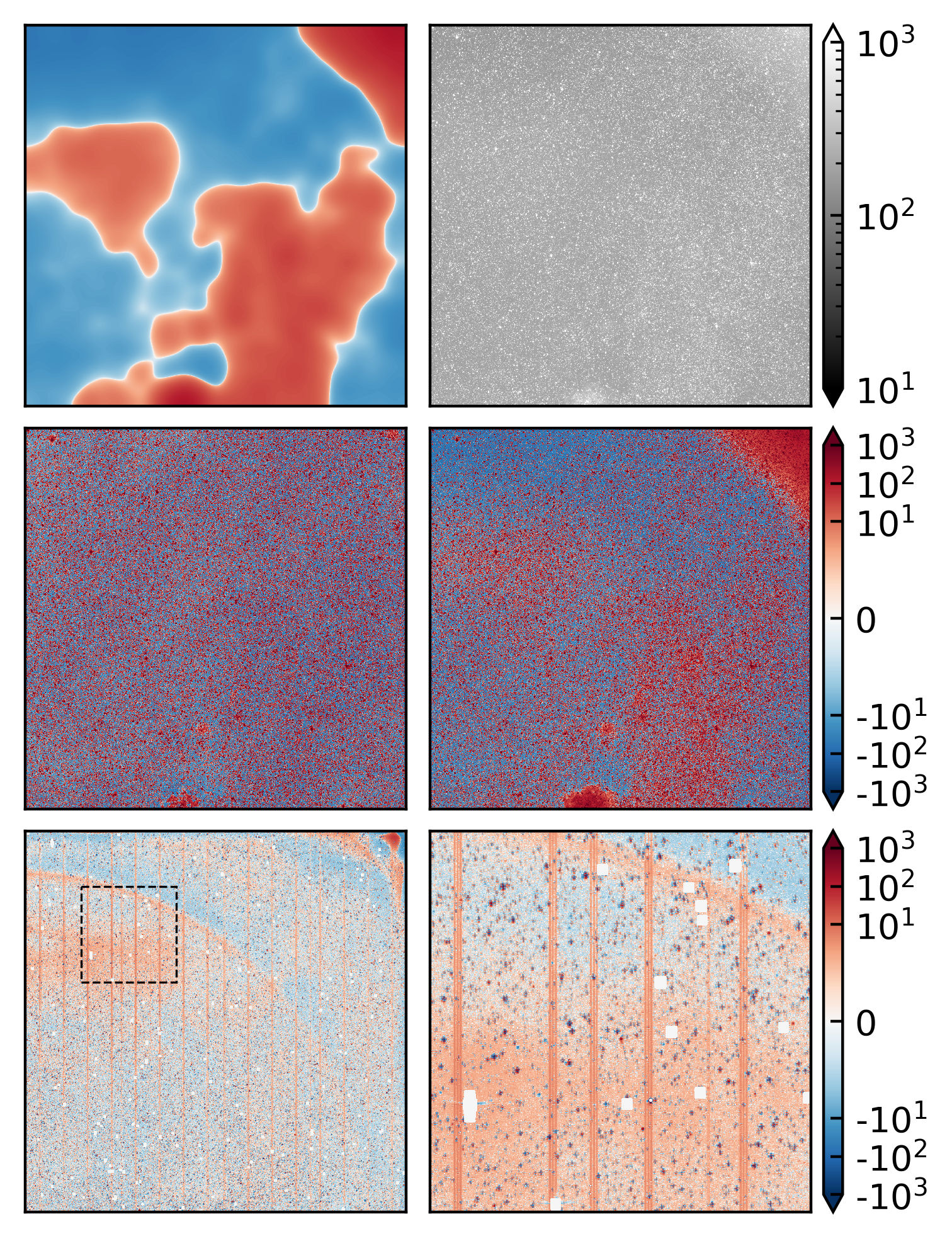}
    \end{center}
    \vspace{-0.7cm}
    \caption{
        Same sector, camera and CCD as Figure~\ref{fig:stages_good},
        but for an image obtained during perigee passage, when
        scattered light from the Earth is prominent.  A number of
        systematic artifacts are present, including vertical
        ``straps'' and small-scale structure in scattered light
        patches.  The quality of astrometric registration in the
        difference image is also worse, leading to larger residuals in
        the lower-right panel.
        \label{fig:stages_bad}
    }
\end{figure*}

\subsection{Overview}

To reduce the TESS images to light curves, we adopted a difference
imaging approach.  The overall method is in the spirit of the
pipelines developed by \citet{Pal_2009},
\citet{huang_high-precision_2015}, \citet{soares-furtado_image_2017},
\citet{oelkers_precision_2018} and \citet{wallace_search_2019}.
Figure~\ref{fig:pipeline} shows a conceptual overview of our pipeline.
Most modules have been developed over the past decade to reduce images
taken by the Hungarian Automated Telescope (HAT) network
\citep{bakos_hat_review_2018}.  The work of \citet{Pal_2009}, embodied
in the \texttt{fitsh} software package, was an especially crucial
component.  The specific high-level framework we used for this
reduction was adapted from a pipeline under development for the HATPI
project (\url{hatpi.org}).  The code is available
online\footnote{\url{github.com/waqasbhatti/cdips-pipeline}, commit
\texttt{7175c48}.}, and the pipeline reference is
\citet{bhatti_cdips-pipeline_2019}.

We begin our processing with the calibrated full frame images produced
by the Science Processing Operations Center at NASA Ames
(\S~\ref{subsec:observations}).  We then perform a collection of
preparatory steps, including source extraction of bright stars,
astrometric verification, and coarse simple aperture photometry of
bright stars (\S~\ref{subsec:preparation}).  Using the information
collected from these initial steps, we select an astrometric reference
frame to which we transform all of the calibrated images.  To
construct a photometric reference frame, we first convolve a subset of
the frames to have identical stellar profiles, and then stack them.
Finally, we subtract each target frame from the photometric reference
frame (\S~\ref{subsec:imagesubtraction}).  We perform aperture
photometry on the subtracted images using positions projected onto the
frame from the Gaia-DR2 source catalog.  The resulting differential
flux measurements are converted to total flux measurements using
photometric information from Gaia-DR2 to determine the total flux of
each source on the reference image.  We detrend the resulting light
curves (\S~\ref{subsec:lcdetrending}).  The resulting white noise and
red noise properties of the light curves, and a few interesting cases
of variability, are explored in \S~\ref{sec:results}.

\subsection{Observations}
\label{subsec:observations}

The TESS spacecraft began science operations on July 25, 2018.  To
keep its cameras pointed opposite the Sun, the spacecraft advances by
$\approx$$28$ degrees east in ecliptic longitude every lunar month.
Data acquired throughout each one-month ``sector'' are downlinked at
spacecraft perigee through the Deep Space Network.  Descriptions of
the spacecraft's design and operations are given by
\citet{ricker_transiting_2015} and \citet{vanderspek_2018}.

For us, the main data product of interest is the calibrated full frame
image (FFI).  Each TESS camera is read out every 2 seconds.  The
resulting pixel values are averaged by the onboard computer into 30
minute exposures. An on-board cosmic ray mitigation algorithm is
applied \citep[][\S 5.1]{vanderspek_2018}. Once transmitted to the
ground, the raw images are calibrated by the Science Processing
Operations Center (SPOC).  The calibration process includes an
overscan, bias, and dark current correction, and also divides out a
flat field.  Details are discussed by \citet{clarke_kepler_2017}, and
the resulting science data products are described by
\citet{tess_data_product_description_2018}.

We perform our processing using the calibrated images, the
corresponding uncertainty images, and the associated headers.  The
spacecraft has four cameras, and each camera has four CCDs.  In the
following analysis, all image-level operations are performed on
individual CCD images, so that at any instant of time there are 16
images that require analysis.

Sectors 1--5 mainly covered portions of the sky away from the galactic
plane.  Consequently, fewer than 2\% of the CDIPS target stars were
observed in the first five TESS sectors.  Although a few interesting
clusters are present in these observations ({\it e.g.}, Blanco~1,
NGC~2516, NGC~1901), for the present work we opted to focus on Sectors
6 and 7, for which there were more stars of interest.  Sector 6
began on December 12, 2018 (space orbit \#19). Sector 7 concluded
on February 1, 2019. Combined, the two sectors
cover galactic longitudes from roughly 200$^\circ$ to 280$^\circ$,
with coverage within $\pm 20^\circ$ of the galactic plane
(Figure~\ref{fig:cdips_targets_positions}).

\subsection{Image preparation \& background removal}
\label{subsec:preparation}

Before we can perform any kind of photometry, a few janitorial tasks
are required.  First, we convert the multi-extension calibrated FITS
image from MAST into a single-extension FITS image, and trim the image
to remove virtual rows and columns using the \texttt{SCIROWS},
\texttt{SCIROWE}, \texttt{SCCSA}, and \texttt{SCCED} header values.

In order to account for the background variations present in some
frames due to scattered light from the Earth and Moon \citep[see][\S
7.3.1--7.3.4]{vanderspek_2018}, we determine and subtract a model of
the large-scale background.  We do this by temporarily masking out
pixels more than $2\sigma$ from the image median, and then pass a
$48\times48$ median box filter over each pixel in the image, with
reflective boundary conditions.  The resulting background estimate has
low-amplitude structure over spatial scales of a few pixels. We then
blur the model image with a gaussian kernel of size 48 pixels, which
produces a smooth background estimate.  These steps also remove
low-level vignetting in the corners of many images, which remains even
after flat-fielding \citep[see][\S 7.3.5]{vanderspek_2018}.  The
results are shown in the upper four panels of
Figures~\ref{fig:stages_good} and~\ref{fig:stages_bad}.  Features with
spatial scales smaller than $\approx$48 pixels remain, but large scale
patterns of scattered light are removed.


After subtracting the background, we mask out saturated pixels using a
fixed saturation level of $8\times10^4$ analog-to-digital units (ADU).
This value was chosen based on the onset of bleeding charge trails in
the images, and is \deleted{slightly}\added{a factor of two} greater than the 
saturation level of
$2\times10^5$ electrons, or about $4\times10^4\,{\rm ADU}$, reported
by \citet{vanderspek_2018}.  As a consequence, we do not analyze stars
brighter than $T\approx 6.5$\deleted{.}\added{, even though the TESS CCID-80 
CCDs conserve charge across bloom trails up to at least
$T\approx 4$ \citet{vanderspek_2018}.}
As described by \citet{Pal_2009}, the
pixel masks are metadata attached to the image file, and are only
applied to the pixel values during the specific image processing steps
in which they are necessary ({\it e.g.}, convolution). We also extend
the masks beyond purely saturated pixels to ``bloomed'' pixels
horizontally and vertically adjacent to the saturated pixels (see
Figure~6 of \citealt{Pal_2009}).

Finally, for frames with the \texttt{DQUALITY} bit-flag corresponding
to the ``momentum dumps'' and ``coarse pointing modes'' described by
\citet{vanderspek_2018}, we omit the entire frame.  This removes on
average a few frames per sector, out of about one thousand. Through
visual inspection, we see that the stars on these frames are extremely
smeared, and are unlikely to produce useful science data.  In
addition, we use the sector-specific data release notes\footnote{\url{
  archive.stsci.edu/tess/tess_drn.html}, accessed \texttt{2019-08-12}} to identify further times
with anomalous spacecraft performance, which we omit from
consideration.  This included three days at the beginning of Sector 6
dedicated to acquiring pixel response function data. There were no
additional gaps in Sector 7.

\subsection{Metadata collection \& WCS verification}
\label{subsec:metadatacollection}

\begin{figure}[!t]
	\gridline{\fig{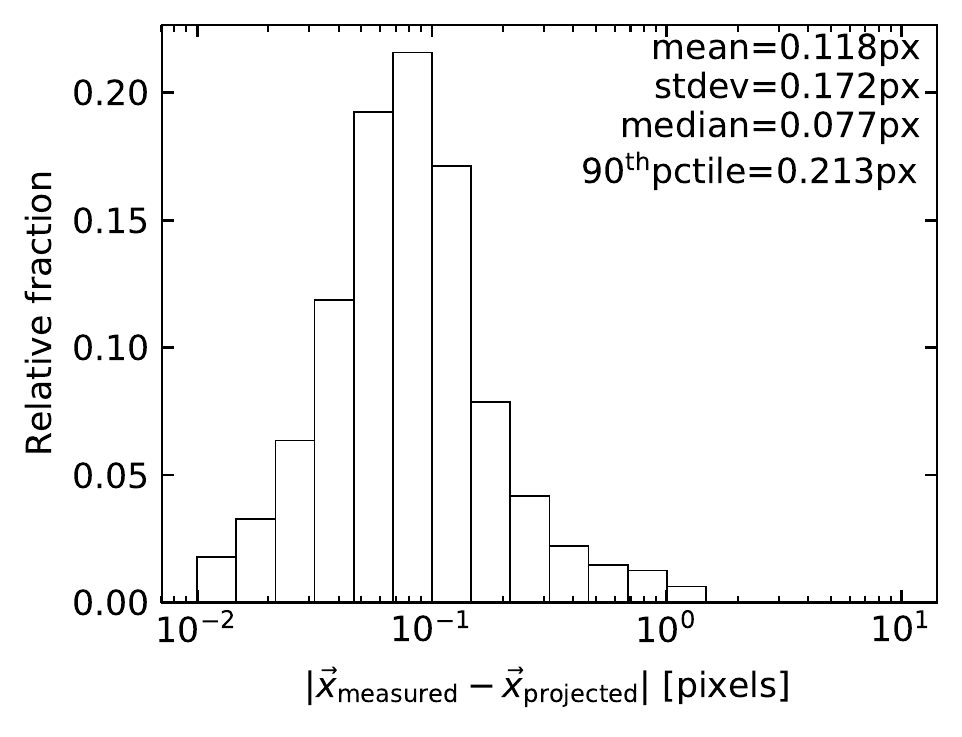}{0.45\textwidth}{}}
	\vspace{-0.8cm}
	\gridline{\fig{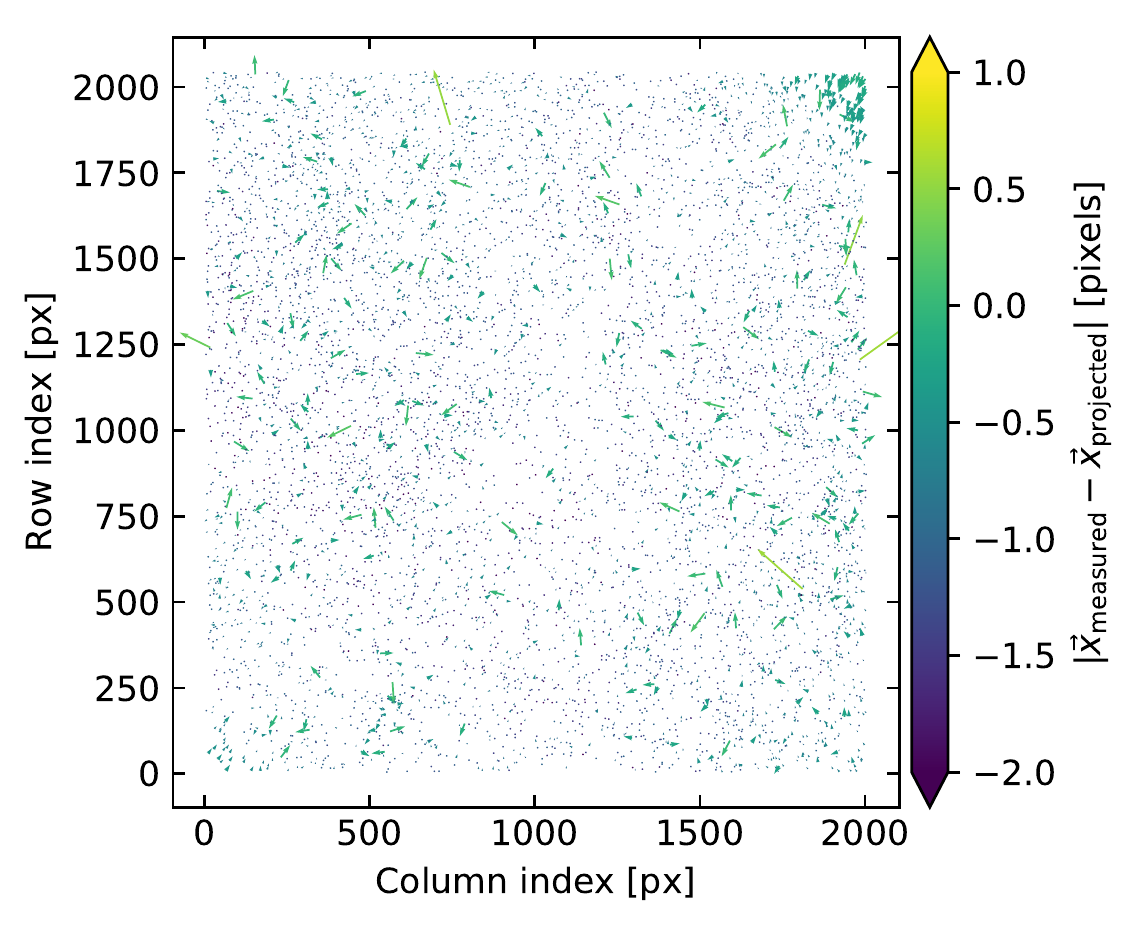}{0.5\textwidth}{}}
	\vspace{-0.8cm}
    \caption{
		{\it Top.} Histogram of astrometric residual. The $x$-axis shows 
		the distance between the measured centroid positions of stars, 
		compared to the predicted positions from the WCS solution.
    {\it Bottom.} Vector plot of astrometric residual. Each arrow is
    the vector from the measured to the projected star position.
    Directions are correct, but lengths are 50 times their true size
    for visual clarity.  The systematic error in the top-right
    corner is a typical problem generic to wide-field astrometry.  The
    frame chosen for this plot is the photometric reference frame used
    for Sector 6, Camera 1, CCD 1; we automatically impose cutoffs on
    the median and $90^{\rm th}$ percentile of the astrometric
    residual in order to ensure similar levels of astrometric
    precision are maintained throughout the reduction.
	}
	\label{fig:astromresid}
\end{figure}

After preparing the images, we perform some initial analysis steps to
produce metadata needed during image subtraction.  

First, we perform source extraction on the thousand or so brightest,
non-saturated stars in each image.  This is done using a
\texttt{fitsh} module, \texttt{fistar}.  We derive centroid positions
for the stars, and simultaneously fit elliptical gaussians to their
profiles, yielding the shape parameters $(s,d,k)$, where the flux $f$
as a function of position $(x,y)$ in the CCD image plane is assumed to
take the form
\begin{align}
  f(x,y) &= B + A \exp \{ -0.5 \times [
    s(\Delta x^2 + \Delta y^2) + \\
    \nonumber
    &d(\Delta x^2 - \Delta y^2) +
    k(2\Delta x \Delta y)
  ]  \},
\end{align}
for $(x_0,y_0)$ the central coordinates of the star, $\Delta x =
x-x_0$, $\Delta y = y - y_0$, $B$ the background level, and $A$ an
arbitrary flux-scaling constant.  For a nearly circular shape profile,
the sharpness $s$ is related to the FWHM as ${\rm FWHM} \approx
2.35s^{-1/2}$ \citep[{\it e.g.},][]{Pal_2009}.  These shape parameters
are later used when selecting an astrometric reference frame
(\S~\ref{subsec:imagesubtraction}).  In agreement with what is obvious
upon visual inspection, this fitting process shows that stars closer
to the center of each camera's field are round, while stars near the
field edges are more elongated.

For the astrometric solution, we use the World Coordinate System (WCS)
and fourth-order Simple Imaging Polynomial (SIP) coefficients derived
by SPOC and included in the FFI headers
\citep[][Sec.~8]{pence_fits_2010}.  We explored the possibility of
using \texttt{astrometry.net} \citep{lang_2010} to derive our own
astrometric solutions for each frame, but found that the astrometric
residual (the mean separation between projected and measured
positions) was consistently a factor of 1.5-2 times higher in our WCS
solutions than in those given by SPOC.  This was perhaps because we
did not develop a robust algorithm to select non-blended stars of
intermediate brightness before measuring their positions.

With the resulting WCS information, we then project a source catalog
onto each frame.  We use the projected positions of the sources to
center the apertures in our photometry, rather than attempting to
measure the positions.  Such ``forced-aperture photometry'' is
preferable to source extraction in the crowded fields that are central
to this work.  The Gaia-DR2 epoch is J2015.5, so even the
fastest-moving stars with proper motions of $\sim$$1\,{\rm
arcsecond}\,{\rm yr}^{-1}$ are still well within one pixel of their
predicted positions in the TESS images.  The projection from catalog
sky-coordinate positions to pixel coordinates is performed using an
analog of the \texttt{wcs-rd2xy} program that performs the standard
matrix algebra \citep{lang_2010}.  The source catalog look-up is
performed using
\texttt{gaia2read}\footnote{\url{github.com/samuelyeewl/gaia2read}, commit \texttt{4b472d}}
\citep{kim_2018_gaia2read}.

For the source catalog itself, we initially planned to photometer all
Gaia-DR2 sources in each field down to a cutoff of $G_{Rp} < 16$.
However, for the galactic plane fields this produced an excessively
large number of sources (millions of stars per
$12^\circ\times12^\circ$ CCD).  We therefore limited our source
catalog for each frame to be a combination of the CDIPS target stars
($G_{Rp} < 16$), and all Gaia-DR2 sources down to $G_{Rp} < 13$.  The
latter set of stars are used for image processing and light curve
detrending.

Figure~\ref{fig:astromresid} displays the residual between the
measured and projected stellar centroid positions for one photometric
reference frame.  The construction of this frame will be described
shortly.  The plot shows that the errors are typically largest in the
corners of the image, where the non-linearity of the focal plane is
most significant, and the corrections required by the SIP coefficients
are largest.  Also, the typical median precision of the WCS solution
is a bit below 0.1 pixels, and its $90^{\rm th}$ percentile is
typically less than 0.3 pixels.  In our reduction, we therefore
require that each frame's median residual and 90th percentile remain
below 0.2 and 0.4 pixels, respectively. If this constraint is not met,
the reduction fails.  This is an essential quality-control check for
any forced-aperture photometry pipeline.

Finally, to collect the metadata needed to select photometric
reference frames, we perform aperture photometry on the bright stars.
This task is performed by using \texttt{fiphot} to sum the counts
inside circular apertures centered on the projected stellar positions.
The pixel weights are equal to the fraction of the pixel that falls
within the circular aperture.  They are unity for pixels entirely
within the aperture, and fractional along the aperture boundary.  The
background levels are measured in annuli surrounding the center of
each aperture.

\subsection{Image subtraction}
\label{subsec:imagesubtraction}

\subsubsection{Synopsis of image subtraction method}

The core operation of ``classical'' image subtraction is to match a
photometric reference image $R$ and a target image $I$ by computing
and applying a convolution kernel.  For ground-based data, this
``match'' typically corrects for differences in seeing or transparency
between the reference and target; for space-based data, the match
might correct for spacecraft jitter, or thermal and corresponding
point-spread function (PSF) variations.  The kernel, once applied to
the high signal-to noise reference, produces a model image, $M_{xy}$,
\begin{equation}
    M_{xy} = (R \otimes K)_{xy} + B_{xy},
    \label{eq:imagemodel}
\end{equation}
where $B_{xy}$ is a component of the model image that allows for
background variations, and $\otimes$ denotes convolution.  Since we
modeled the background separately (\S~\ref{subsec:preparation}), we
set $B_{xy}=0$.  The convolution kernel $K$ is typically decomposed
onto a basis,
\begin{equation}
K = \sum_i c_i K_i,
\end{equation}
where the coefficients $c_i$ are found by minimizing
\begin{equation}
    \chi^2 = \sum_{xy} \left( \frac{I_{xy} - M_{xy}}{\sigma_{xy}} \right)^2,
    \label{eq:chisq_conv}
\end{equation}
for $\sigma_{xy}$ the uncertainty in the target image pixel
values.  Photometry is then performed on the difference image
$D_{xy}$, where $D_{xy} = I_{xy} - M_{xy}$.  For the present
reduction, the uncertainty in each target image pixel was taken to be
a constant.	

The general procedure described above was first proposed by
\citet{Alard_Lupton_1998}.  It was reviewed and clarified by
\citet{miller_optimal_2008}.  The choice of how to decompose the
kernel was further explored by \citet{bramich_new_2008}, who showed
that using a linear combination of delta functions (also called a
``discrete kernel'') had advantages compared to a basis of gaussians.
We perform the convolution using \texttt{ficonv}, and opt for the
implementation of Bramich's method (see \citealt{Pal_2009} \S~2.8).
The lower panels of Figure~\ref{fig:stages_good} show the procedure
working well, and producing a ``clean'' difference image.
Figure~\ref{fig:stages_bad} shows what happens for an image taken when
scattered light from the Earth causes the model image to be a poor fit
to the target image.

\subsubsection{Astrometric registration}

To make the above high-level picture work, we need to select two
``reference frames'': (1) the astrometric reference frame; and (2) the
photometric reference frame.

To choose the astrometric reference frame, we search for frames with
compact, round stars (big $s$, small $d$ and $k$ values).  We also
require that the frame have a low background level, as measured in
annuli around the bright stars selected in
\S~\ref{subsec:preparation}.  Finally, the astrometric reference frame
needs to have a large number of detected sources (though the variance
between TESS images was rather small).  We sort the images using these
metrics, and then select the astrometric reference frame from
successive intersections of each sorted list.

We then compute and apply a spatial transformation 
to each calibrated frame in order to match the astrometric reference.
This transformation~--~a combination of rotation, dilation, and
translation~--~typically moves stars by less than a pixel, since the
TESS spacecraft pointing is quite stable.  We calculate the
transformation using the measured source positions found in
~\S~\ref{subsec:metadatacollection}, and the symmetric triangle
point-matching scheme described by \citet[][~\S~2.5.2]{Pal_2009}.
This step is achieved using the \texttt{fitsh} tools \texttt{grmatch}
and \texttt{grtrans}.  To help ensure the precision of the
transformation, we require the ``unitarity'' $\Lambda$
\citep[][~Eq.~54]{Pal_2009}, which characterizes the degree of
distortion in the transformation matrix, to be below 0.01.  To
mitigate possible photometric errors incurred during this step, we
also use the flux-conserving interpolation scheme described by
\citet{Pal_2009}, which is necessary because polynomial interpolation
schemes do not conserve stellar flux. 

\subsubsection{Photometric reference frame construction \& reference flux measurement}
\label{subsubsec:photref}

The second required reference frame is the photometric reference
frame, which is used both to calculate the convolution kernel, and to
obtain a reference flux for each star.  To make it, we first choose 50
images with low background measurements (measured for each frame from
the annuli around bright stars), and only consider frames with a
relatively large number of detected bright objects.  We then convolve
these candidate photometric reference frames to the frame with the
lowest background measurement, and construct the photometric reference
as the median image across the 50 frame stack. 

Measuring the reference flux for each star is a non-trivial operation.
First, we perform forced simple aperture photometry on the photometric
reference frame to measure the flux for each source.  The local
background is estimated in annuli, with neighboring stars masked out
during the background measurement.  If we were to stop here, {\it it
would be a mistake}.  The reference fluxes for faint stars would be
overestimated, due to crowding.  The relative amplitude of photometric
signals for faint stars would correspondingly be biased to lower
values.

To avoid this problem, after performing simple aperture photometry on
the reference frame, we fitted a line between the TESS T-band magnitude
of the bright stars, and their measured fluxes.  The TESS $T$-band magnitudes were
calculated using the Gaia-DR2 magnitudes of each star, and
Equation~1 of \citet{stassun_TIC8_2019}.  We
then used the known catalog magnitudes for all the stars on
each image to predict the expected reference
flux for each star.  This accounts for crowding down to Gaia's
resolution limit of $\approx$1$''$, rather than the TESS limit of
$\approx$20$''$.

The final instrumental flux values $f$ we report are given by
\citep[][Equation~83]{Pal_2009} 
\begin{align}
  f &=  f_{\rm reference} + f_{\rm subtracted} \\
  &=
  g \left(T_{\rm cat} \right)
  +
  \frac{1}{|| K ||_1^2} \sum_{x,y} D_{xy} (w \otimes K)_{xy}.
  \label{eq:ism_flux_measurement}
\end{align}
The function $g$ takes as input the target star's catalog magnitude
$T_{\rm cat}$, and returns the reference flux.  Its coefficients are
found independently for each aperture.  The difference image $D$ is
equal to $I -  (R\otimes K)$, where as in
Equation~\ref{eq:imagemodel}, $I$ is the target image transformed to
the astrometric reference, $R$ is the photometric reference, and $K$
is the convolution kernel.  The weights $w$ from the circular aperture
mask are included in the convolution.  The norm $|| K ||_1$ is defined
by \citet{Pal_2009}~Equation~81.

\subsection{Choice of convolution kernel}
\label{subsec:tuneconvkernel}

To solve for the coefficients $c_i$ of the convolution kernel, a few
further assumptions are necessary.  The procedure implemented in
\texttt{ficonv} is to subdivide the image, and within each grid
element find the brightest non-saturated star. These isolated
``stamp'' stars are then used to solve for the coefficients of the
kernel, by minimizing Equation~\ref{eq:chisq_conv} over the sum of all
stamps.  For the kernel basis, we use a linear combination of delta
functions with a flux scaling term
(\citealt{soares-furtado_image_2017} Section 3.3.1 gives the
equations).  In this model, spatial variations of the PSF across the
image are captured by weighting each basis component with spatial
polynomials up to a cut-off order.

This kernel model has three parameters that must be specified, but are
not automatically optimized by the procedure: (1) the box-size; (2)
the maximum order of the polynomial weighting the delta function
terms; (3) the maximum order of the polynomial weighting the flux
scaling.  We performed a grid-search to tune these parameters, in
which our ``loss functions'' were the light curve standard deviation
(RMS) as a function of magnitude, and the recovered SNR of transits
from the catalog of known TOIs (TESS Objects of Interest; N.~Guerrero,
in preparation).

In the first dimension, we varied the kernel size between a box of
$3\times3$ pixels and $11\times11$ pixels.  Increasing the kernel
box-size from a $3\times3$ box to a $7\times7$ box led to about a 50\%
lower light curve RMS for bright stars, and no difference for faint
stars.  The largest kernels, of $(11\times11)$ pixels, returned
slightly lower signal-to-noise for recovered transits than kernels of
intermediate size.  We settled on a kernel box-size of $(7\times7)$
pixels, which is $\approx$2 times larger than the typical TESS FWHM at
field center. 

In the other two dimensions, we varied the spatial polynomial orders
weighting the kernel's individual pixels between first and fifth
order.  We did the same for the polynomial weights of the ``identity''
pixel.  Varying the polynomial orders between 1 and 4 did not produce
large differences.  The fifth order polynomials retrieved transits
with $\approx10\%$ worse SNR compared to lower order polynomials.  We
therefore adopted a second order polynomial weight in both terms.

Averaging over all TOIs present in the camera we used for these
experiments, we found that different choices of kernel parameters
produced variations of $\lesssim 12\%$ in the retrieved transit SNR.
For computational expediency, we therefore chose a single $(7\times
7)$ kernel with second-order spatial polynomial weights in the basis
functions for the remainder of our reduction.

With a kernel selected, and the convolution and subtraction performed,
we calculated the instrumental fluxes for each frame per
Equation~\ref{eq:ism_flux_measurement}.  We did this with three
different aperture sizes: for this work, circles of radii 1 pixel, 1.5
pixels, and 2.25 pixels.  These sizes were chosen to roughly span the
range of optimal aperture sizes reported by
\citet{Sullivan_et_al_2015}.  Finally, to convert from a list of flux
measurements for each source on a frame to light curves, we used the
\texttt{fitsh} transposition tool \texttt{grcollect}.

\subsection{Light curve detrending}
\label{subsec:lcdetrending}

\begin{figure*}[!t]
	\gridline{
    \fig{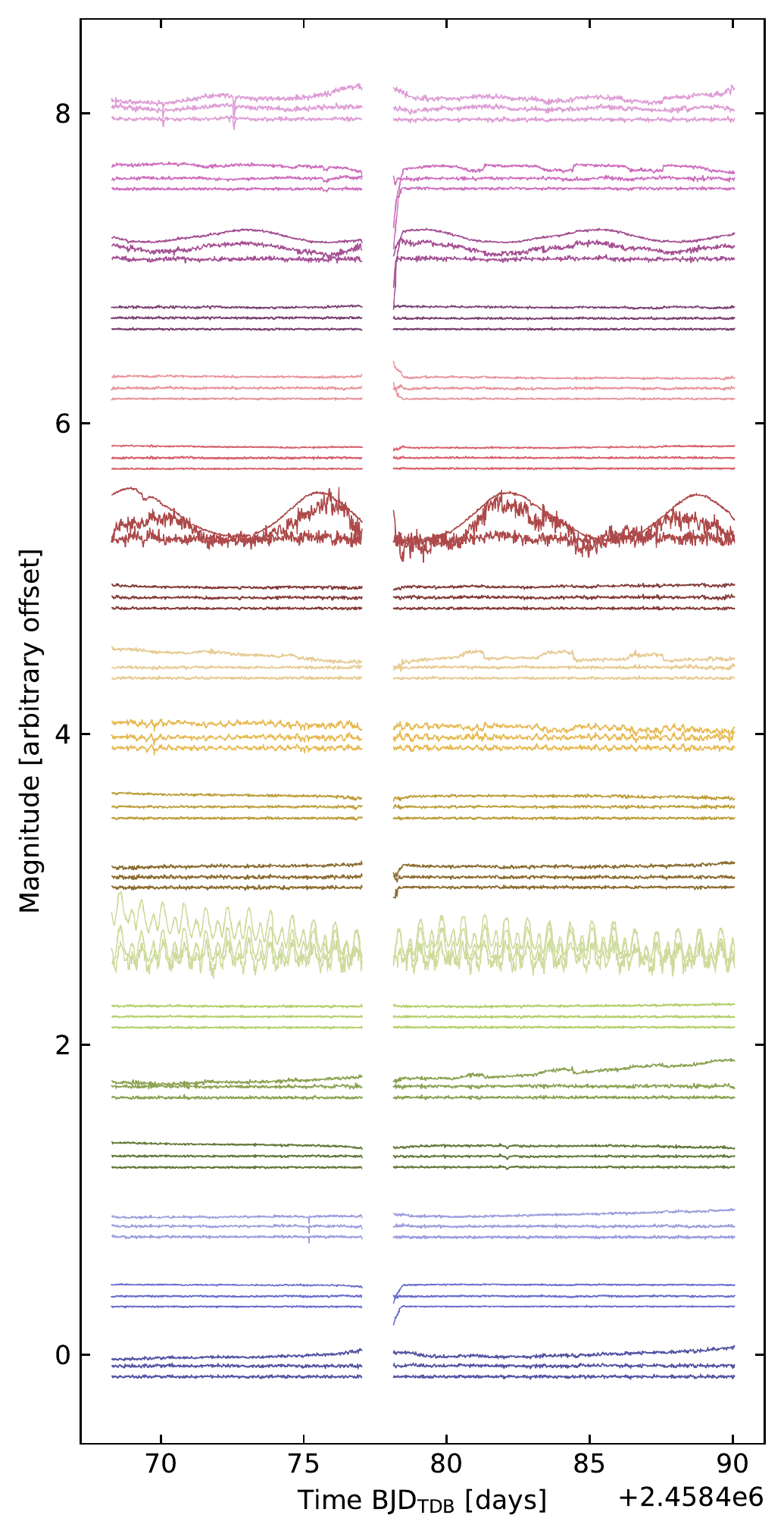}{0.5\textwidth}{}
    \fig{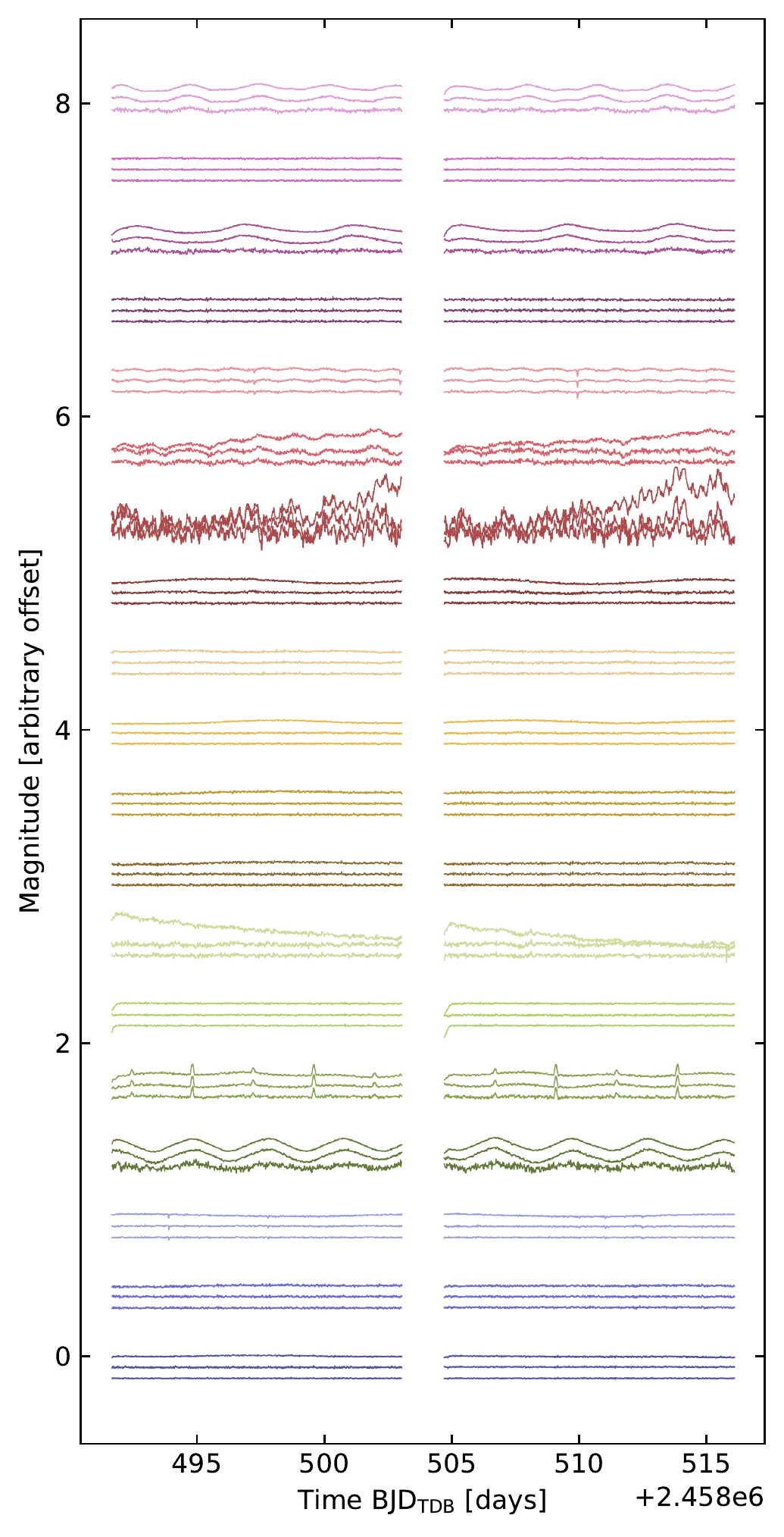}{0.5\textwidth}{}
  }
	\vspace{-0.85cm}
	\caption{
    Twenty randomly selected light curves drawn from the same sector,
    camera, and CCD, for CDIPS target stars with $T$-band magnitudes
    between 13 and 14.  For each star, we show the raw light curve
    (top), PCA-detrended light curve (middle), and TFA-detrended light
    curve (bottom).  The sector, camera, and CCD numbers are 6, 1, 4
    (left) and 7, 3, 2 (right).  A number of systematic trends are
    shared across raw light curves ({\it e.g.}, the periodic $\sim$3 day
    ``chopping'' seen in the left plot).  In the PCA light curves,
    stellar rotation signals are usually preserved, but not always ({\it
    e.g.}, left panel, seventh from the top).  TFA filters out almost
    all long-term trends, or else heavily distorts them ({\it e.g.},
    right panel, fourth from the bottom)
	\label{fig:lc_systematics_dtr}
	}
\end{figure*}

\begin{figure}[!t]
	\begin{center}
		\leavevmode
		\includegraphics[width=0.45\textwidth]{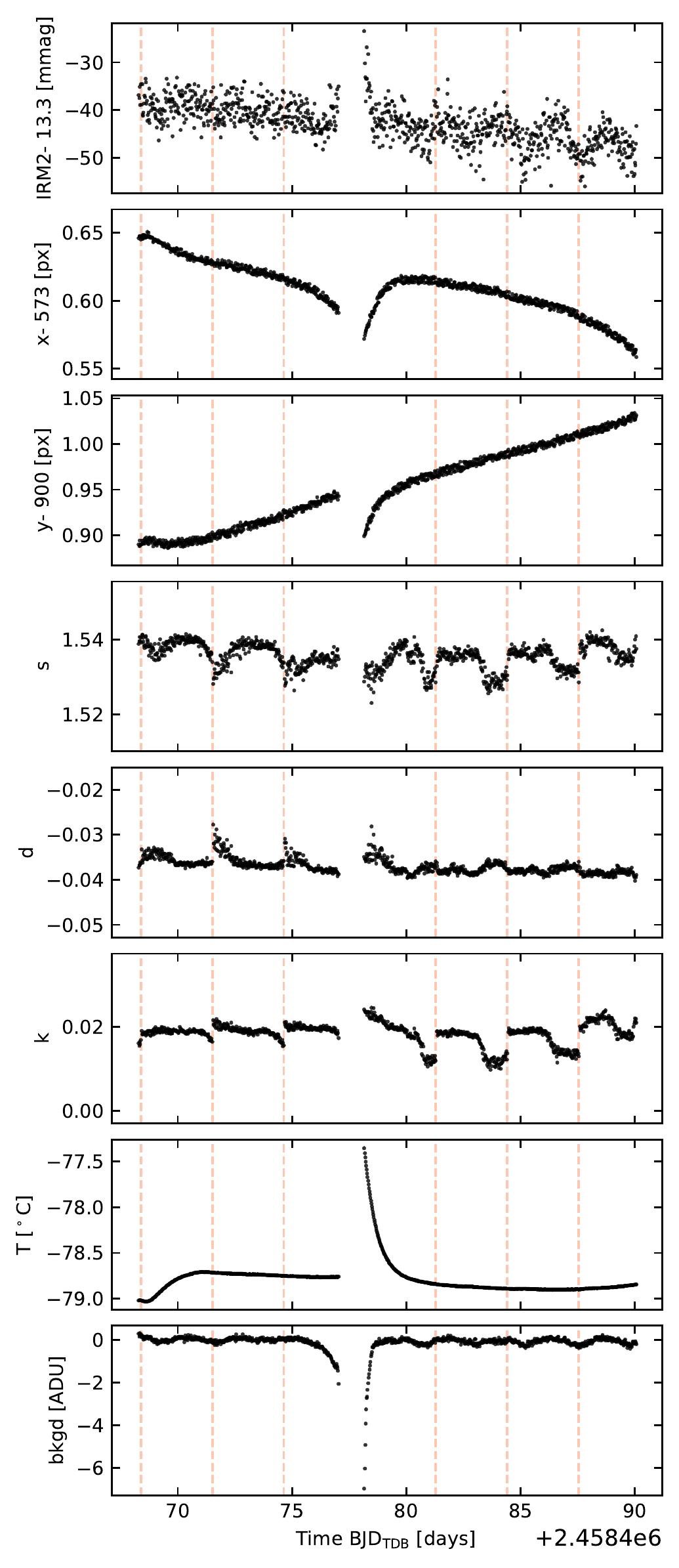}
	\end{center}
	\vspace{-0.5cm}
	\caption{
    The variability in flux is sometimes correlated with variability
    in ``external'' parameters, shown here for a representative star
    over two orbits.  {\it Top}: Instrumental raw magnitude (with a
    particular aperture size), as a function of time.  Continuing in
    order are $x$ and $y$ centroid positions as functions of time,
    the $(s,d,k)$ PSF shape parameters, the CCD temperature,  and the
    measured background value.  Differential aberration affects the
    centroid position over the span of each orbit.  Momentum dumps are
    marked with vertical dashed lines, and affect the measured shapes
    of stars.
		\label{fig:external_parameter_timeseries}
	}
\end{figure}

The preceding steps produce light curves that include both
instrumental systematics as well as astrophysical variability.
Figure~\ref{fig:lc_systematics_dtr} shows twenty stars of comparable
brightness randomly selected from two CCDs.  Stars that are far apart
on the same CCD often share similar changes in flux.  In other words,
the instrumental systematics seem to dominate.  This problem is
generic in wide-field photometric datasets, including the WASP,
Kepler, and HAT surveys
\citep{Pollacco_2006,borucki_kepler_2010,bakos_hat_review_2018}.  To
remove the systematic variability, we adopted two different
approaches: {\it (i)} the trend-filtering algorithm (TFA,
\citealt{kovacs_trend_2005}), and {\it (ii)} a principal component
analysis (PCA, see {\it e.g.}, \citealt{ivezic_statistics_2014} for a
review).

However, a number of other approaches to the problem were possible. To
encourage future improvements, we describe the possibility of
decorrelating against external parameters
(\S~\ref{subsubsec:external}), and also different available approaches
to ensemble detrending (\S~\ref{subsubsec:ensemble}), before
explaining our adopted implementation.

\subsubsection{Decorrelating against external parameters}
\label{subsubsec:external}

Often, ensemble trends of stellar magnitude with CCD position,
sub-pixel position, catalog magnitude, and color are present in
datasets.  One detrending step that can be valuable is to fit and
subtract a linear combination of these trends as they appear across
many light curves \citep[{\it e.g.},][\S~5.5]{zhang_precision_2016}.

A separate step for each light curve can then be to fit out linear
correlations of stellar magnitude with ``external parameters'' (EPD,
\citealt[][]{bakos_2010,huang_high-precision_2015}).  For ground-based
data these parameters might include zenith angle, or changing PSF
shape.  For TESS data, they might include CCD temperature, or the
angles of the Moon and Earth relative to each camera's boresight.
They might also include the standard deviation of the spacecraft
\deleted{quartnerion}\added{quaternion} time-series
\citep{vanderburg_hr858_2019}.  Some example ``external'' parameters
that we include with our light curves are shown as functions of time
in Figure~\ref{fig:external_parameter_timeseries}.

We explored the possibility of fitting linear models of flux as
functions of {\it e.g.,} temperature, shape parameters, and centroid
positions to each light curve.  We also briefly explored non-linear
model fitting using $N$-dimensional B-splines to fit the flux,
centroid positions, and temperatures simultaneously
\citep{dierckx_curve_1996}.  The linear models typically underfit the
light curves, particularly during the large shifts that happen as the
spacecraft nears perigee.  The non-linear models showed some promise,
but often seemed to overfit stellar variability signals.  Given these
complications, for the time being we omitted the step of
``detrending'' as a function of external parameters. To enable further
exploration of the issue, we included all the necessary vectors of
{\it e.g.}, centroid positions, temperatures, and shape parameters in
our reported light curves.

\subsubsection{Ensemble detrending}
\label{subsubsec:ensemble}

The parameters that capture systematic trends are often poorly known.
In such cases, an effective model of the systematics comes from
constructing a set of basis vectors that empirically captures trends
common to many stars.  Each target light curve is then assumed to be a
linear combination of the trend vectors.

The well-known algorithms, TFA, Sys-Rem, PDC-MAP, and ARC2, all take
slightly different approaches to constructing this set of basis
vectors, as well as to solving for the weights to assign each linear
component
\citep{kovacs_trend_2005,tamuz_correcting_2005,smith_pdc_2012,aigrain_robust_2017}.
TFA selects individual ``template stars'' as basis vectors, and
equally weights each template when solving for the coefficients via
linear least squares.  PDC-MAP computes ``co-trending basis vectors''
(CBVs) by applying singular value decomposition to the light curves
that show the strongest mutual correlation.  It solves for the
coefficients through a two-step procedure.  The first step is to
calculate the coefficients through linear least squares.  The
least-squares coefficients are then used to construct a prior over
plausible coefficient values, which is subsequently used to recompute
the maximum likelihood coefficients for each star.  This latter step
reduces overfitting for stars with variability not present in the set
of CBVs.

We opted to use two different detrending approaches, each aimed at a
different use case.  For transit-search related science, we used TFA,
as implemented in \texttt{VARTOOLS}
\citep{kovacs_trend_2005,Hartman_Bakos_2016}.  For stellar
astrophysics related work, we used a simple variant of PCA, as
implemented in \texttt{scikit-learn} \citep{sklearn_2011}.  For
self-consistency, we describe each method and its implementation in
the following paragraphs.

\paragraph{TFA detrending}

The idea of TFA is as follows.  Suppose we have $M$ ``template
stars'', which are a subsample of stars that represent all types of
systematics across the dataset.  Each template star has a light curve
with $N$ data points.  Denote the template time-series $X_j(i)$, where
$j={1,\ldots,M}$ and $i={1,\ldots,N}$ is the time index.  We then want
to find periodic signals in a target time-series $Y(i)$.  This is done
by defining a filter function
\begin{equation}
  F(i) = \sum_{j=1}^{M} c_j X_j(i),
\end{equation}
for which the coefficients $c_j$ are found by minimizing
\begin{equation}
  \mathcal{D} = \sum_{i=1}^{N} \left[ Y(i) - A(i) - F(i) \right]^2.
  \label{eq:tfa_to_minimize}
\end{equation}
When trying to find periodic signals, $A(i)$ represents our prior
knowledge of the light curve's shape.  Initially, this prior is simply
that stars on average maintain a constant brightness:
\begin{equation}
  A(i) = \langle Y \rangle = \frac{1}{N} \sum_{i=1}^{N} Y(i) = {\rm const.}
\end{equation}
If a signal is eventually found, for instance using the box-least
squares method \citep{kovacs_box-fitting_2002}, this detrending
process must then be repeated while accounting for our updated
knowledge about the light curve's shape.

Some implementation notes follow.  We selected template stars in two
stages.  In the first stage, we fitted a parabola in the RMS-magnitude
plane, and discarded stars more than $2\sigma$ away from the
prediction of the fit.  We also required that these initial candidate
stars have intermediate brightness ($8.5 > T > 13$), and have a
relatively large number of time-series data points.  We excluded
templates within 20 pixels of any given target star.  We then
performed an initial iteration of TFA, on only the candidate template
stars.  We inspected the resulting detrended light curves for residual
structure by computing a Lomb-Scargle periodogram.  If the
maximum-power peak had a false alarm probability below 0.1\%, we
excluded the star from the list of candidate template stars, on the
basis of its presumed periodic variability.  We then randomly selected
at most 200 template stars from the remaining non-variable candidates.
The choice of number of template stars was discussed by
\citet{kovacs_trend_2005}, and is another free parameter in the broad
problem of light curve production.  This choice is analogous to the
issue of how many cotrending basis vectors to choose in PDC-MAP or
ARC2 \citep{aigrain_robust_2017}.  While the number of template stars
can be optimized by constructing and minimizing a BIC-like quantity, a
little overfitting is acceptable for our pupose of finding planetary
transits.  For other applications, {\it e.g.,} stellar rotation period
searches, it is almost certainly preferable to adopt a less forceful
detrending approach.

\paragraph{PCA detrending}

To remove the largest systematic trends with minimal overfitting of
{\it e.g.}, stellar rotation signals, we adopted a simple variant of
PCA.  A similar approach was taken by \citet{feinstein_eleanor_2019}
in \texttt{eleanor}.  We derived the principal components for each CCD
using the 200 template stars previously selected for TFA.  We then
modelled each target light curve as a linear combination of a subset
of these principal components, and determined the weights via linear
least squares. Both steps were performed using \texttt{scikit-learn}
\citep{sklearn_2011}.

To determine the number of principal components at which to truncate
the model, we performed a cross-validation analysis.  Again, this was
achieved using \texttt{scikit-learn}.  As is typically the case when
the noise variance is different for each ``feature'' (each point in
time), we found that $k$-folding cross-validation of the PCA
components gave a cross-validation score that monotonically increased
with an increasing number of components\footnote{See for example
\url{https://scikit-learn.org/stable/auto_examples/decomposition/plot_pca_vs_fa_model_selection.html}, accessed \texttt{2019-08-06}.}.  We found that if we instead performed a factor
analysis, the cross-validation score was typically maximized with
anywhere from $10$ to $15$ components.  This agreed with a visual
analysis of the number of PCA components past which overfitting began.
We therefore used the maximum cross-validation score from the factor
analysis as the number of principal components to use in each light
curve.  This number is documented in the \texttt{FITS} header of each
light curve.

While the PCA detrending should help remove the most egregious
systematic artifacts, it remains possible that PCA can distort
signals, and even inject correlated noise into a light curve ({\it
e.g.}, Figure~\ref{fig:lc_systematics_dtr}).  For any object of
interest, it is worth inspecting the raw, PCA, and TFA light curves to
ensure that the variability of interest is not adversely affected by
the detrending procedures.  If it is, alternative approaches may be
necessary.

\section{Results}
\label{sec:results}

\begin{figure*}[!t]
	\begin{center}
		\leavevmode
		\includegraphics[width=0.9\textwidth]{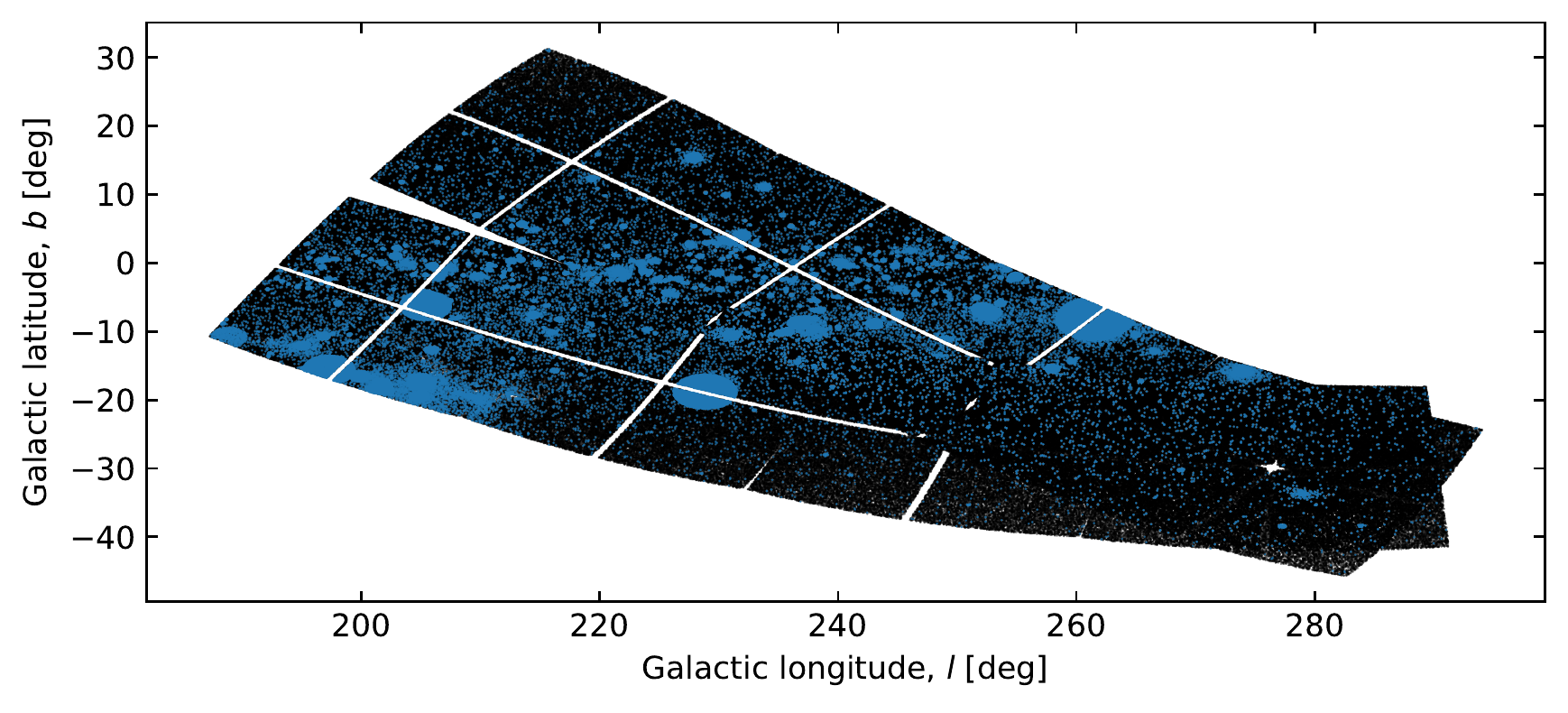}
	\end{center}
	\vspace{-0.7cm}
	\caption{
    Positions of light curves from TESS Sectors 6 and 7 in galactic
    coordinates.  Black: $G_{\rm Rp}<13$ field stars.  Blue: $G_{\rm
    Rp}<16$ target stars.  Target stars are mostly near the galactic
    plane. The data for Sectors 6 and 7 cover about one-sixth of the
    galactic plane.
		\label{fig:lcgalactic}
	}
\end{figure*}

\begin{figure}[!t]
	\begin{center}
		\leavevmode
		\includegraphics[width=0.45\textwidth]{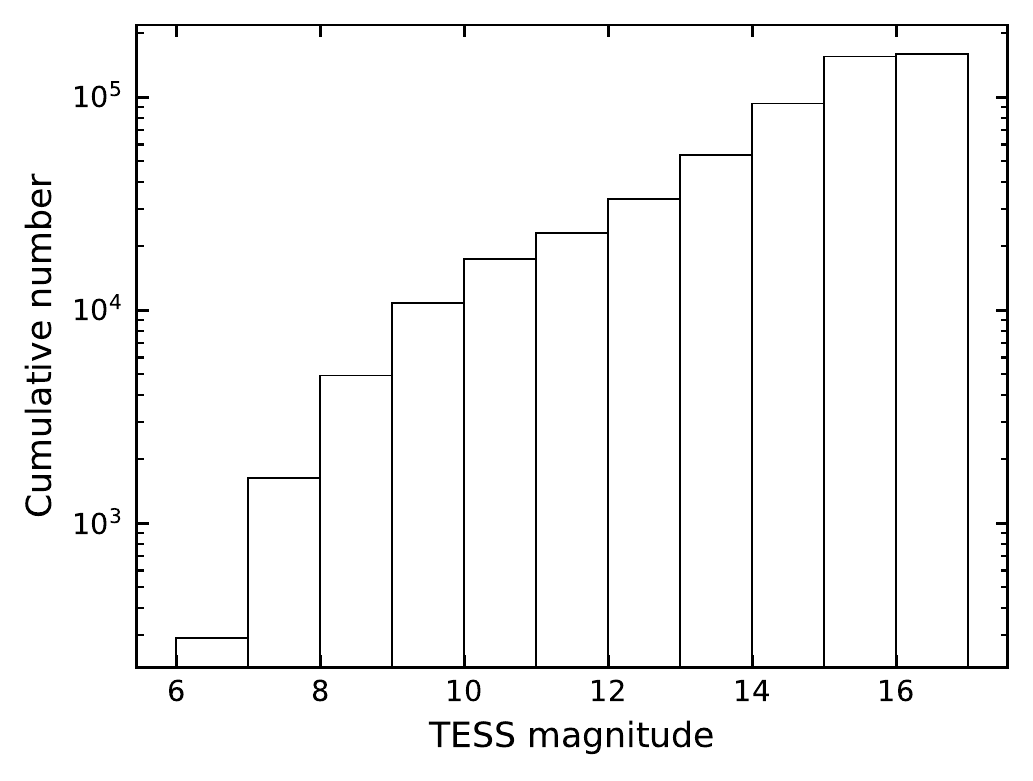}
	\end{center}
	\vspace{-0.5cm}
	\caption{
    Cumulative number of CDIPS light curves a function of TESS
    $T$-band magnitude.  Light curves were made for the target stars
    (Figure~\ref{fig:cdips_targets}) that were observed in Sectors 6
    or 7.
		\label{fig:cdf_T_mag}
	}
\end{figure}

\begin{figure}[!ht]
	\gridline{\fig{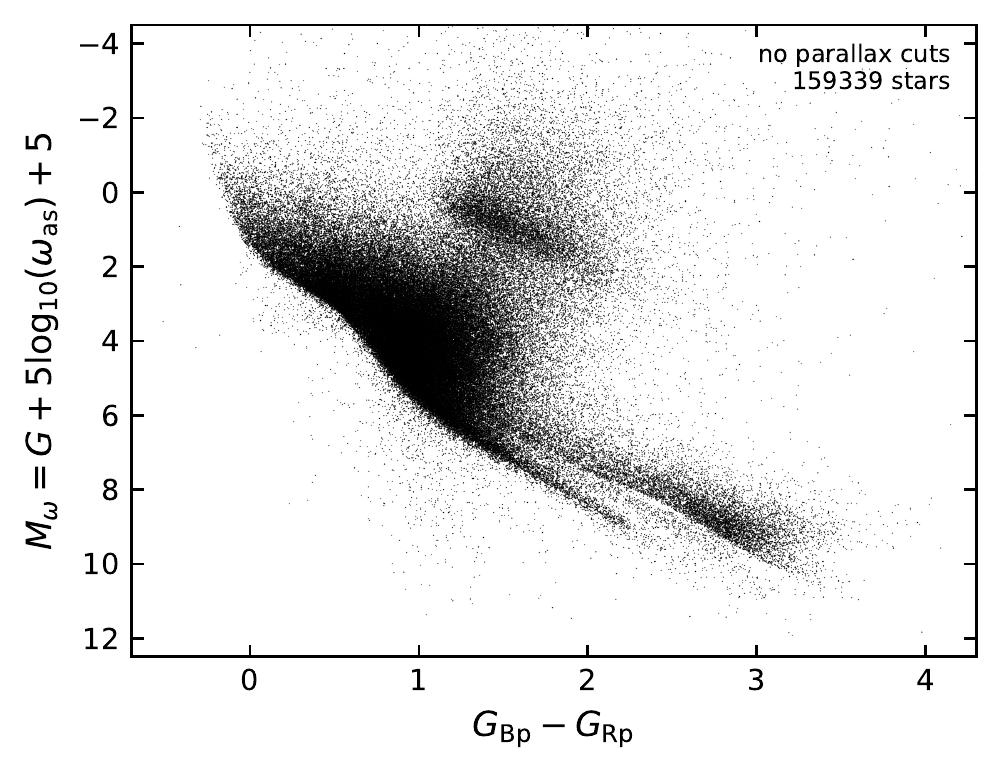}{0.45\textwidth}{}}
	\vspace{-1.1cm}
	\gridline{\fig{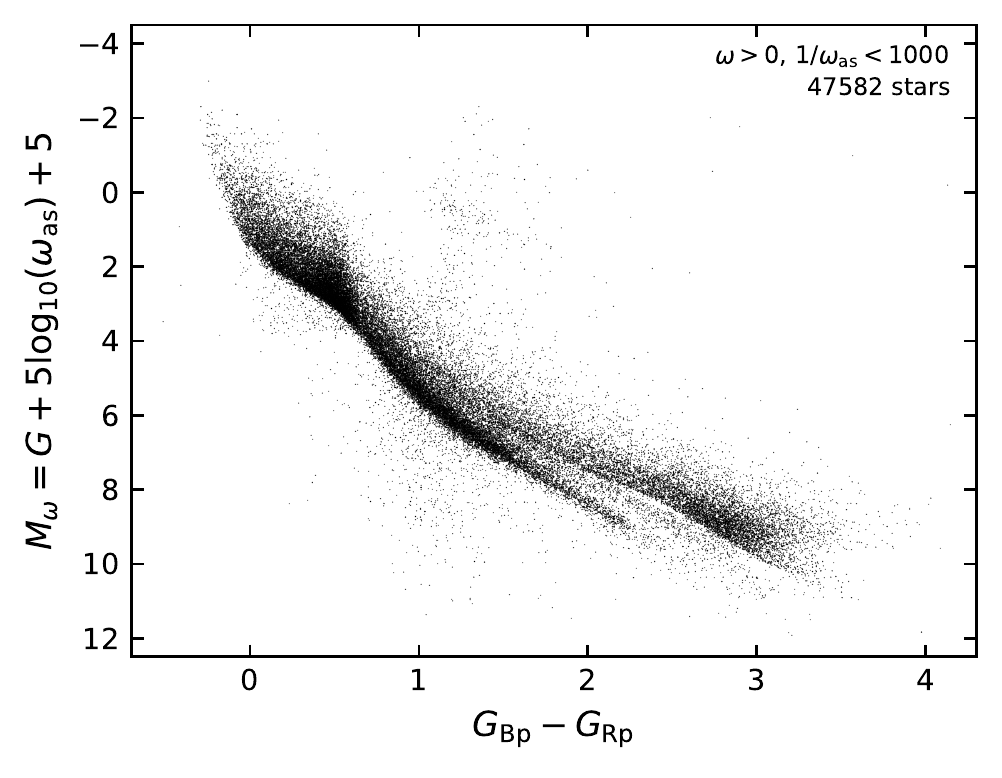}{0.45\textwidth}{}}
	\vspace{-0.9cm}
	\caption{
    {\it Top.} HR diagram of CDIPS stars on silicon in this data
    release.  {\it Bottom.} HR diagram of close CDIPS stars on
    silicon. The wedge separating the pre-MS sample from the MS stars
    was discussed by \citet{zari_3d_2018}, who introduced it in order
    to avoid contamination by photometric binaries.
	}
	\label{fig:hrd}
\end{figure}

\begin{figure}[!t]
	\gridline{\fig{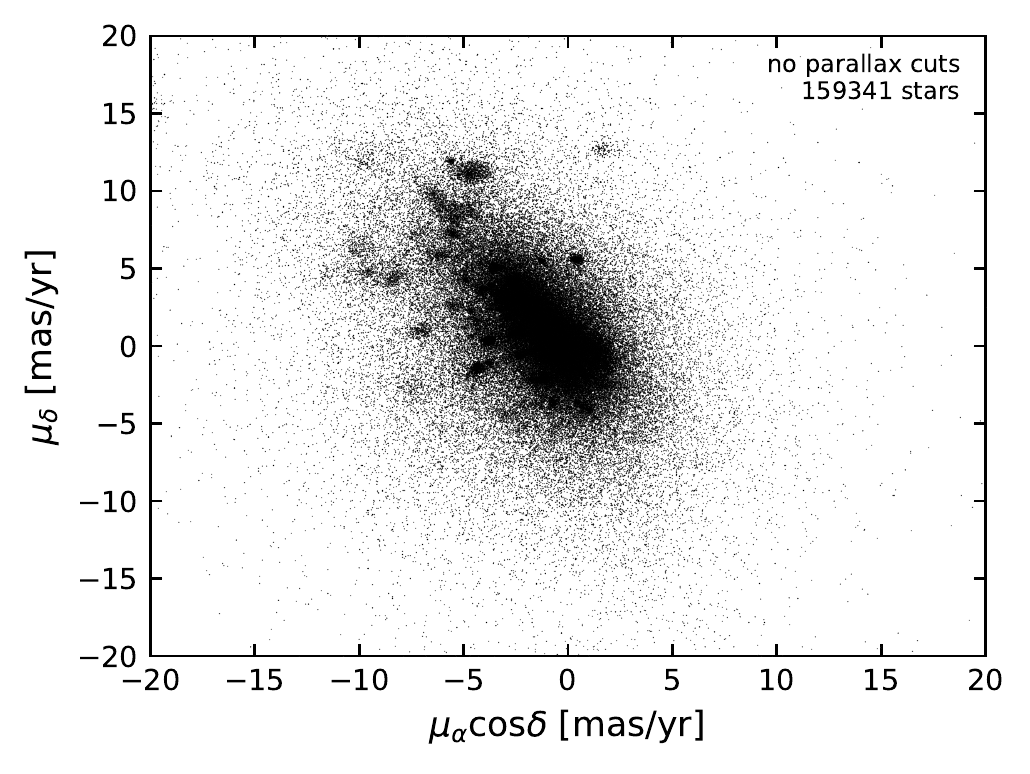}{0.45\textwidth}{}}
	\vspace{-0.9cm}
	\gridline{\fig{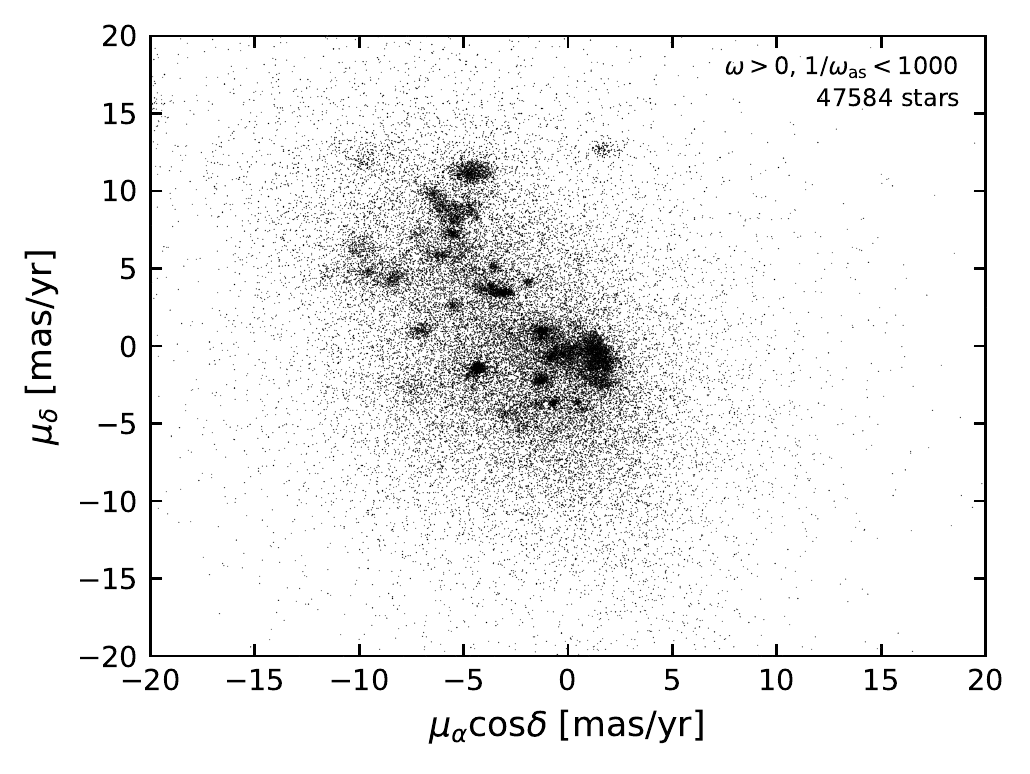}{0.45\textwidth}{}}
	\vspace{-0.9cm}
	\caption{
		{\it Top.} Proper motions of CDIPS stars on silicon in this
		data release.  Many of the stars in the central ``blob'' are possible
		field-contaminants.
		{\it Bottom.} Proper motions of close CDIPS stars
		on silicon.
	}
	\label{fig:propermotions}
\end{figure}

\begin{figure*}[!t]
	\begin{center}
		\leavevmode
		\includegraphics[width=0.9\textwidth]{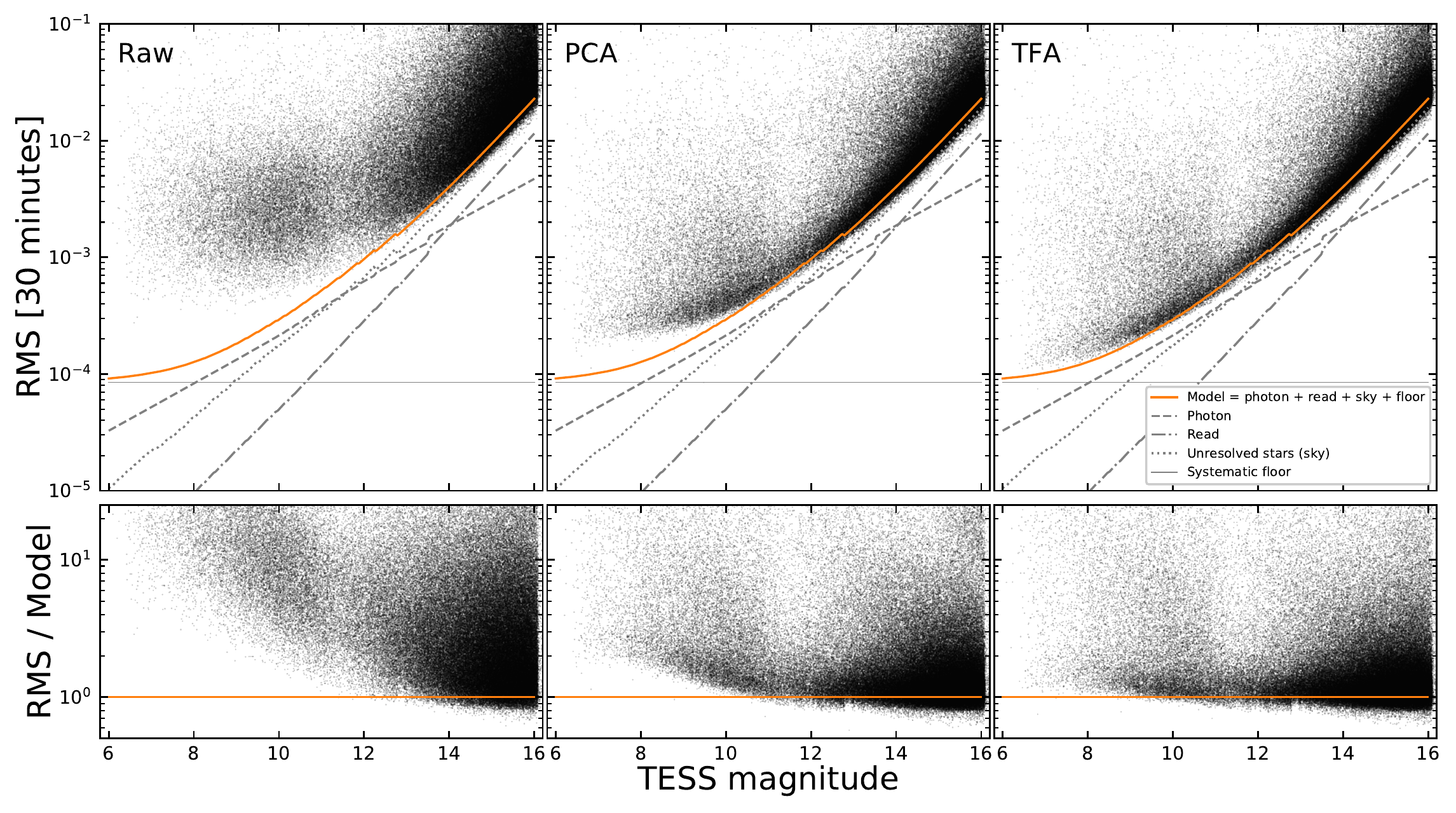}
	\end{center}
	\vspace{-0.7cm}
	\caption{
    Standard deviation of CDIPS light curves as a function of catalog
    TESS-band magnitude.  Left-to-right are raw, PCA-detrended, and
    TFA-detrended light curves. Black points correspond to the minimum
    RMS across the three available aperture sizes.  The model (orange
    and gray lines) assumes aperture sizes calculated by
    \citet{Sullivan_et_al_2015}, and the effective area from
    \citet{vanderspek_2018}.  The noise from unresolved background
    stars (dotted gray line) is a function of galactic latitude, and
    dominates over zodiacal light for faint stars near the galactic
    plane; the line shown assumes a sight-line towards the center of
    Sector 6, Camera 1 (further details are in
    \S~\ref{subsubsec:rmsvsmag}).
		\label{fig:rms_vs_mag}
	}
\end{figure*}

\begin{figure*}[!t]
	\begin{center}
		\leavevmode
		\includegraphics[width=0.85\textwidth]{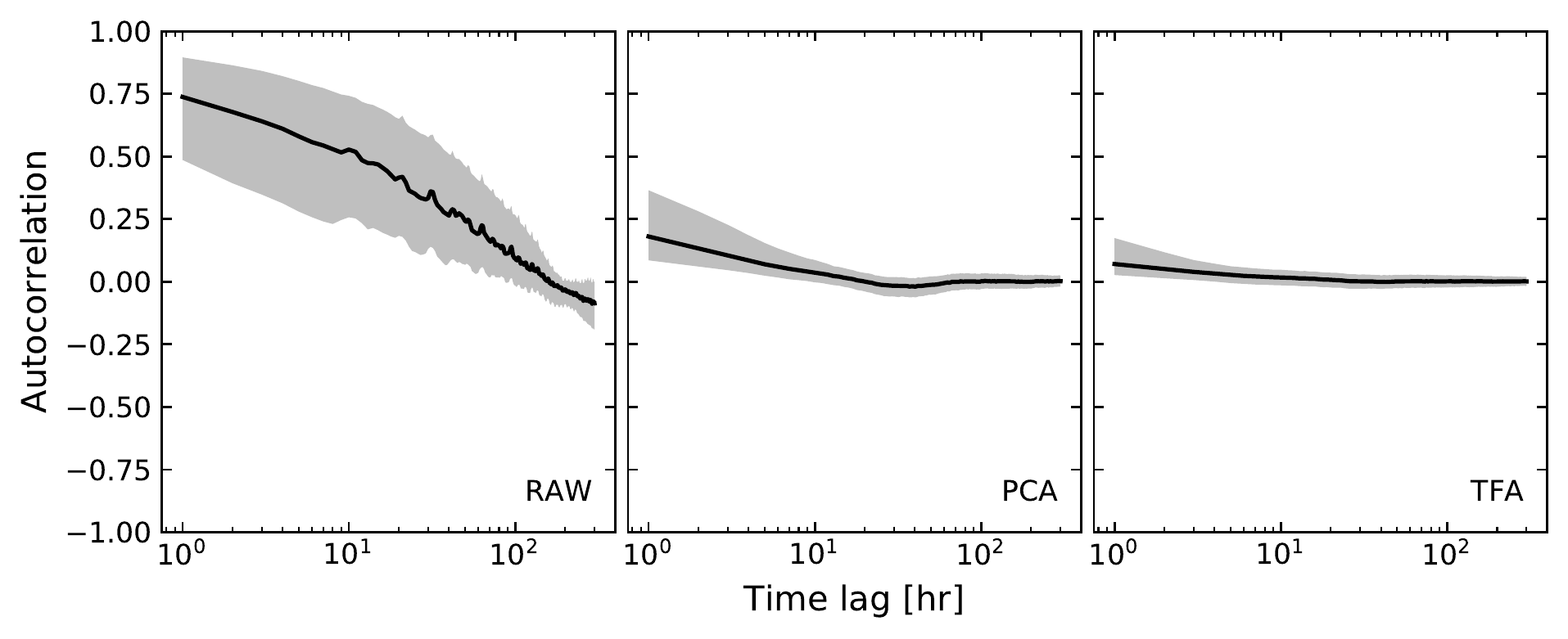}
	\end{center}
	\vspace{-0.7cm}
  \caption{
    Average autocorrelations of $10^4$ randomly selected light curves
    from each sector.  The medians at time lags spaced by one hour are
    shown for raw light curves (column name \texttt{IRM}) in gray, and
    for TFA-detrended light curves in black.  The gray bands display
    the 25$^{\rm th}$ to 75$^{\rm th}$ percentile range for each type
    of light curve.
  \label{fig:avg_acf}
	}
\end{figure*}

\begin{figure*}[!t]
	\begin{center}
		\leavevmode
		\includegraphics[width=1\textwidth]{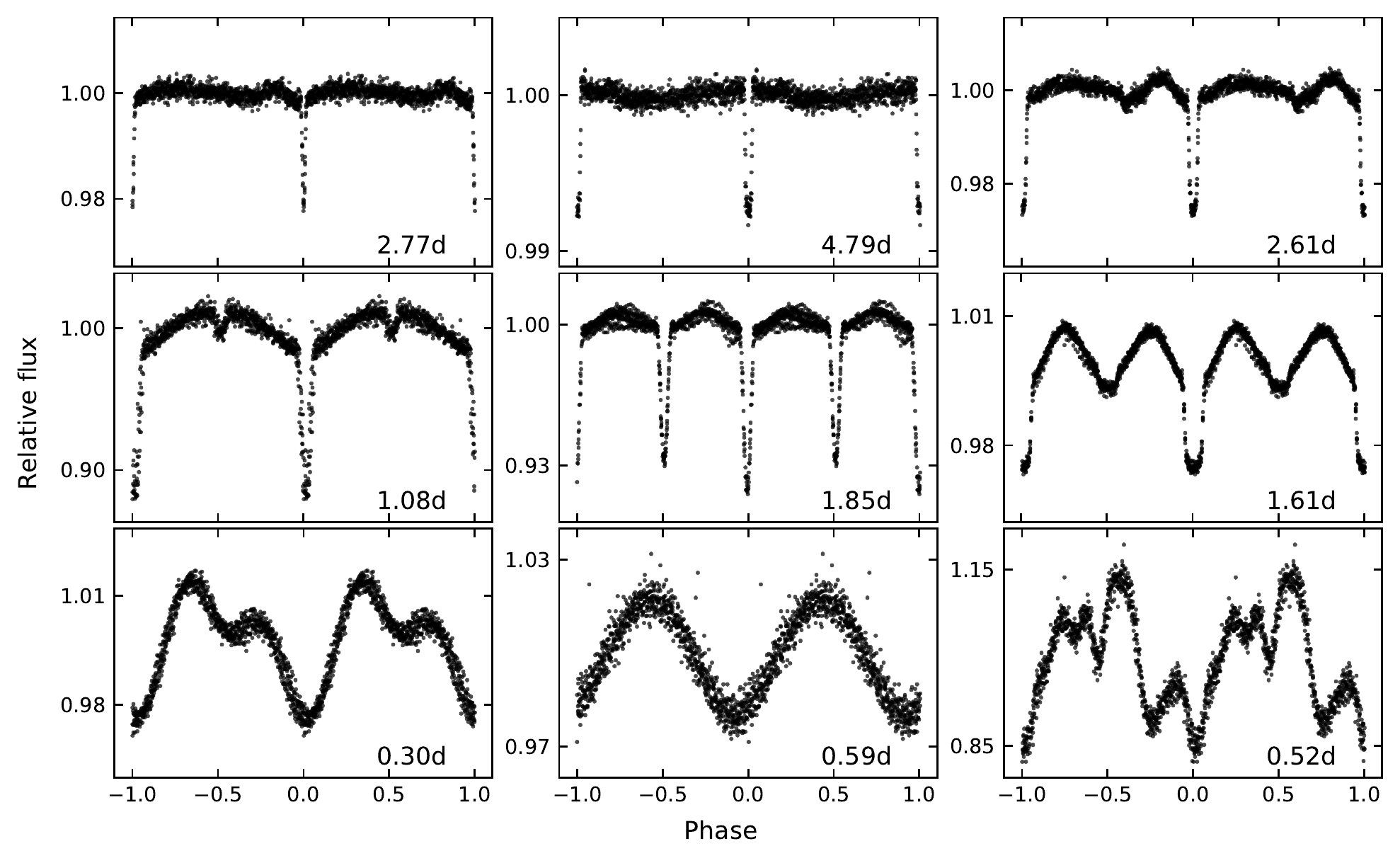}
	\end{center}
	\vspace{-0.8cm}
	\caption{
		Phased PCA light curves for a few objects of interest.
		{\it Top row, left to right}:
    TOI~496 in Messier~48; TOI~625 orbiting a late A dwarf; and an
    eccentric EB in Bochum~5.
		{\it Middle row, left to right}:
    A detached EB in a 80 Myr old cluster; V684~Mon (a 10 Myr old
    detached EB); and a semi-detached EB in NGC~2548.
		{\it Bottom row, left to right}:
    Spotted rotator in NGC~2184; V468~Ori, a flaring M dwarf in
    Messier~42 with a strong rotation signal; Rapidly rotating 30 Myr
    old M dwarf, similar to those described by
    \citet{zhan_complex_2019}.  Gaia source identifiers are given in
    the text.
	\label{fig:quilt}
	}
\end{figure*}

\subsection{Light curve statistics}
\label{subsec:lcstatistics}

\subsubsection{Stellar properties}

Figure~\ref{fig:lcgalactic} shows the on-sky locations of the light
curves from Sectors 6 and 7.  Black points are the field stars for
which we performed photometry in order to construct basis vector sets
for TFA and PCA.  Blue points are the \numberlcs target stars for
which light curves are available at \datasetlink.  Most of the target
stars are close to the galactic plane, and are spatially clustered.
Gaps between CCD chips are also visible.

Figure~\ref{fig:cdf_T_mag} shows the cumulative distribution of TESS
$T$-band magnitudes for the target stars.  Though most of the targets
are faint, $\approx3\times10^4$ are brighter than $T=13$. The brighter
targets will be more amenable to detailed spectroscopic observations,
should the need arise.

An HR diagram for the entire sample of CDIPS stars on silicon is shown
in Figure~\ref{fig:hrd} (top).  The sub-sample of stars with measured
positive parallaxes and naive distances less than $1\,{\rm kpc}$ is
also shown (bottom).  About one-third of the stars are in this latter
sample.  Of the close stars, a relatively large fraction come from
\citet{zari_3d_2018}, and are either on the PMS or upper
main-sequence.  The latter set of OBA dwarf stars, while ``younger''
than the typical field dwarf, are the least interesting subset of our
target sample from the perspective of age analyses.  In the entire
sample (Figure~\ref{fig:hrd} top), a much larger fraction of stars are
sub-giants, red giants, and helium-burning red-clump stars.

Finally, Figure~\ref{fig:propermotions} shows the proper motions of
the entire and close samples of stars on silicon.  Each clump
signifies a different star cluster. The large overdensity in the top
panel is composed mainly of field star contaminants. It has two
subcomponents, due to the two different directions in the Galaxy being
observed in Sectors 6 and 7.

\subsubsection{Cluster membership provenance}

\paragraph{Sector 6}

In Sector 6, \sVInumberlcs light curves of candidate cluster stars
were made. The provenance of the target star for these sources is
\citet{dias_proper_2014} for 40\% of the sources; \citet{zari_3d_2018}
for 11\% of sources from their upper main-sequence table and 7\% of
sources from their PMS table; \citet{Kharchenko_et_al_2013} for 18\%
of sources; \citet{cantat-gaudin_gaia_2018} for 14\% of sources; and
more than two catalogs for the remaining 10\% of sources.

Table~\ref{tbl:s6_lcs} lists the clusters in Sector 6 with the largest
numbers of light curves. The top four clusters are all moving groups
or stellar associations from \citet{dias_proper_2014}: Platais~5,
Platais~6, Mamajek~3, and Collinder~70.  These membership claims
should be regarded with skepticism.  For Platais~6,
\citet{Kharchenko_et_al_2013} claimed only about 400 probable members
(1$\sigma$) to exist within the angular radius of the cluster.
Mamajek~3 (= 32$\,$Ori) similarly has only about 50 confirmed members
\citep{bell_32ori_2017}.  Some rich open clusters in the Sector 6
field with both many light curves and more reliable membership lists
include Trumpler~5, Collinder~69 (the $\lambda$ Orionis cluster), and
NGC~2287.

\paragraph{Sector 7}

In Sector 7, \sVIInumberlcs light curves of candidate cluster stars
were made.  The provenance of the claimed cluster origin of these
sources is \citet{dias_proper_2014} for 28\% of the sources;
\citet{Kharchenko_et_al_2013} for 24\% of sources;
\citet{zari_3d_2018} for 8\% of sources from their upper main-sequence
table and 3\% of sources from their PMS table;
\citet{cantat-gaudin_gaia_2018} for 22\% of sources, and more than two
catalogs for the remaining 15\% of sources.

Table~\ref{tbl:s7_lcs} shows the clusters from Sector 7 with the
largest number of light curves.  Many rich open clusters, including
NGC~2437, 2477, 2546, and 2451 are in the field.  The cluster listed
with the most light curves is the Collinder 173 association, chiefly
due to the \citet{dias_proper_2014} memberships.
\citet{Kharchenko_et_al_2013} note Collinder 173 as being coincident
with the Vela OB2 association, and hence they flag it as a
``duplicate''.  Along with Sco-Cen and the Orion star forming complex,
the Vela OB2 association is one of the largest young associations
within 500~pc \citep{zari_3d_2018}.  Crucially though,
\citet{cantat-gaudin_velaOB2_2019} recently studied the region using
Gaia DR2 astrometry, and found that they could subdivide the complex
into seven distinct stellar populations, with a total of $\sim$11,000
members.  None of these populations map one-to-one to the classic
labels of ``Collinder 173'' or ``Vela OB2''~--~so the
\citet{cantat-gaudin_velaOB2_2019} catalog is likely a more reliable
source of information for this particular region of the sky.

\subsubsection{Light curve noise properties}
\label{subsubsec:rmsvsmag}

\paragraph{Observed RMS vs magnitude.}

Figure~\ref{fig:rms_vs_mag} shows the standard deviation of the
TFA-detrended light curves as a function of the catalog $T$-band
magnitude for CDIPS target stars.  For the $y$-axis of this plot, we
have taken
\begin{equation}
  {\rm RMS} = \left[
    \frac{1}{N-M-1}
    \sum_{i=1}^{N} (f_i - \bar{f})^2
  \right]^{1/2},
\end{equation}
where $f_i$ is the value of the flux at the $i^{\rm th}$ point in the
time-series, $\bar{f}$ is the mean flux, $N$ is the number of points
in the time-series, and $M$ is the number of template light curves
(principal components) used during TFA (PCA) detrending.  The
correction in the denominator is for the overfitting inherent in any
ensemble detrending method.

The observed RMS (black points) follows the usual shape, with source
photon noise dominating from $T=9$ to $T=12$, beyond which the onset
of the ``sky'' background changes the overall slope of the curve to be
more steep.  For the brightest stars ($T\lesssim 9$), a ``systematic
floor'' of 60$\,$ppm$\,$hr$^{1/2}$ was part of the mission's error
budget \citep{ricker_transiting_2015}, but has not been observed in
early reports of the photometric performance of various aperture
photometry pipelines ({\it e.g.,} the SPOC pipeline
\citealt{jenkins_spoc_2010}, the MIT-QLP \citealt{huang_tess_2018},
and \texttt{eleanor} \citealt{feinstein_eleanor_2019}).  The fact that
our light curves for the brightest stars are above this purported
``floor'', rather than below it, suggests that our image subtraction
techniques could be introducing some small degree of noise to the
light curves of the brightest stars.  It is also true however that our
largest aperture contains only about 16 pixels, which is sub-optimal
for stars brighter than $T\approx9$
\citep[][Figure~14]{Sullivan_et_al_2015}.  Since the brightest
stars are not the focus of the present work, we leave this issue
unaddressed for the time being.

An essential feature of our RMS diagram is that faint stars in crowded
regions are not spuriously driven below theoretical limits.  In
aperture photometry pipelines, a star in a crowded region has its
reference flux overestimated relative to the true number of photons it
contributes (due to contamination from neighbor stars).  As a result,
changes in relative flux in the faint star's light curve are
underestimated, and its RMS is driven low \citep[{\it e.g.}, the faint
end of][Figure~5]{feinstein_eleanor_2019}.  Our method to work around
this issue -- using the catalog magnitudes to predict the reference
flux values, and measuring deviations from these reference fluxes on
the subtracted images -- seems to be performing as intended.

\paragraph{Expected RMS vs magnitude.}

The noise model shown in Figure~\ref{fig:rms_vs_mag} is quite similar
to that of \citet{Sullivan_et_al_2015}, save for two changes.  The
first change is that the effective area of the telescope is updated to
be $86.6\,{\rm cm}^2$, per the measurements by
\citet{vanderspek_2018}.

The second change is that we have explicitly included the estimated
noise contribution from unresolved faint stars.  The brightness of the
diffuse sky is dominated by different sources at different
wavelengths.  For instance, the CMB is most important in the
microwave, and thermal radiation from dust grains in the solar system
(zodiacal light) is dominant in the far infra-red
\citep{leinert_1997_1998}.  In the TESS-band, both zodiacal light and
faint stars can play a role, depending on the line of sight under
consideration. The zodiacal light is brightest near the ecliptic
plane, and the faint star background is brightest near the galactic
plane.  When performing pre-launch noise estimates,
\citet{winn_background_2013} estimated the photon-counts from each
component.  His zodiacal light model was presented by
\citet{Sullivan_et_al_2015}, but the faint star component was not
emphasized since the Sullivan simulations were performed for fields
away from the galactic plane.

The diffuse sky model we have used for Figure~\ref{fig:rms_vs_mag} is
adopted explicitly because most of our target stars are near the
galactic plane.  Stars are judged to be ``unresolved'' and part of the
background if their surface density exceeds the angular resolution of
the telescope.  TESS has an angular resolution of $\Delta \theta \sim
1'$, set by a combination of the pixel size as well as the typical
stellar FWHM.  Sources with sky surface density exceeding $\Delta
\theta^{-2}$ therefore contribute to the background.

The relevant quantity needed to calculate the integrated photon counts
from faint sources is  $N(<m,l,b)$ --- the number of stars per square
arcsecond brighter than magnitude $m$, along a line of sight with
galactic longitude and latitude $(l,b)$.  To calculate this surface
density, \citet{winn_background_2013} queried the Besan\c con model
\citep{robin_synthetic_2003}  along a grid of galactic sight-lines,
and then converted to $I$-band surface brightnesses.  Fitting a smooth
function to the results, \citet{winn_background_2013} found
\begin{equation}
  I\ {\rm mag\ arcsec}^{-2} =
      a_0 + a_1 \left(\frac{|b|}{40^\circ}\right)
      + a_2 \left(\frac{|l|}{180^\circ}\right)^{a_3},
\end{equation}
where the galactic longitude $l$ is measured from $-180^\circ$ to
$180^\circ$, and the empirical coefficients were found to be $a_0 =
18.9733$, $a_1=8.833$, $a_2=4.007$, and $a_3=0.805$.  The model is
{\it very approximate}.  It is sensitive to the threshold used to
select ``unresolved'' stars, and likely no more accurate than 0.5 mag
on average.  In regions with exceptionally high extinction ({\it
e.g.}, star forming regions) it is expected to
\deleted{systematic}\added{systematically}
underestimate the background brightness by an even larger degree.
Nonetheless, this model for the diffuse sky background seems to agree
reasonably well with the observed trend of standard deviation versus
stellar magnitude.

\paragraph{ACF statistics}

Beyond the white noise properties of the light curves, the red noise
properties are also important.  In Figure~\ref{fig:avg_acf} we show
the average autocorrelation of the raw, PCA, and TFA light curves.
The raw light curves have substantial red noise, and their average
autocorrelation across many time lags is positive.  The PCA and TFA
light curves are much closer to white noise~--~the average
autocorrelation between any two points in these light curves is close
to zero.  However, the average PCA and TFA light curves are not {\it
completely} uncorrelated.  At time lags of a few hours or less, there
is some excess power.  This suggests that additional detrending may be
necessary to maximize the effectiveness of planet searches, or the
discovery of any other signals via matched-filter techniques.

\subsection{Exploring the variability}
\label{subsec:identifying_variability}

The light curves provide an opportunity to study many types of
variables, including pulsating stars, rotators, eclipsing binaries,
and transiting planets.  A few hand-picked examples are shown in
Figure~\ref{fig:quilt}.
The sources\footnote{
	Top-left to bottom-right:
	3064530810048196352, 
	3080104185367102592, 
	3027361888196408832,
	3125738423345810048,
	3326715714242517248,
	3064487241899832832,
	3024952755135530496,
  3209428644243836928,
  3214130293403123712 
} were identified by calculating Lomb-Scargle and transit-least
squares periodograms for a subset of the light curves, and inspecting
the peaks with the greatest power
\citep{lomb_1976,scargle_studies_1982,kovacs_box-fitting_2002,vanderplas_periodograms_2015,hippke_TLS_2019}.
We calculated the periodograms using \texttt{astrobase}
\citep{bhatti_astrobase_2018}.  Our search was cursory~--~a detailed
search for transiting exoplanets will be the subject of future work.
To make the figure, we used the \texttt{PCA2} light curve for each star.
To remove outliers, we omitted data points within 6 hours of the
beginning or end of each orbit.  No additional detrending was
performed.

The top row of Figure~\ref{fig:quilt} shows two TESS objects of
interest, and an eccentric eclipsing binary.  TOI~496 is a member of
Messier~48 (NGC~2548,
\citealt{gaia_collaboration_gaia_2018,cantat-gaudin_gaia_2018}), a
$\sim$500 Myr old open cluster \citep{Kharchenko_et_al_2013}.  The
phase variations suggest that it is an eclipsing binary.  TOI~625 is a
potential hot Jupiter orbiting an upper main-sequence star, and was
included as a CDIPS target through the \citet{zari_3d_2018} catalog.
The eccentric EB was claimed by \citet{dias_proper_2014} to be a
member of the $\sim$50 Myr Bochum~5 cluster.  This EB is also notable
because its TFA light curve heavily whitens the out-of-eclipse signal,
leaving an eclipse signal that could be mistaken for a planet
candidate.  The relevant lesson for anyone using the light curves is
to verify that detrending does not introduce or remove signals that
alter the interpretation of the system.

The middle row of Figure~\ref{fig:quilt} shows a few more interesting
eclipsing binaries.  On the left is a detached EB exhibiting large
out-of-eclipse modulations in a $\sim$80~Myr open cluster (``vdBergh
85'', \citealt{Kharchenko_et_al_2013}).  In the middle we have V684
Mon, a well-known $\sim$10~Myr old detached EB in the very young
star-forming region NGC~2264.  Finally on the right, we have a
semi-detached EB in Messier~48.

The bottom row of Figure~\ref{fig:quilt} gives a few examples of
rotational variables.  On the left, a spotted rotator in the
$\sim$400 Myr NGC~2184 \citep{cantat-gaudin_gaia_2018}.  In the
middle we have V468~Ori, a flaring M dwarf in Messier~42 with a strong
rotation signal.
Finally on the right is a $\sim$30 Myr old M dwarf in ASCC~19
with complex rotational modulation.
The star was also flagged by \citet{zari_3d_2018} as a pre-MS star.
Similar modulations were described by \citet{zhan_complex_2019} for
rapidly rotating M dwarfs in other moving groups, and were suggested
to be caused by star spot occulations behind a protostellar disk.

Overall, our periodogram search showed that 7\% of the light curves
have Lomb-Scargle peak false alarm probabilities below $10^{-30}$.
Even more stars are variable at lower levels of signifiance.  This implies
that thousands of variable stars with known ages should
be identifiable from the data at their current level of preparation.

\section{Discussion \& Conclusion}
\label{sec:conclusion}

In this study, we collected an all-sky sample of about one million
stars brighter than $16^{\rm th}$ magnitude in $G_{\rm Rp}$.  82\% of
these target stars are candidate ``cluster'' members, where we use the
generic ``cluster'' to refer to open clusters, stellar associations,
and moving groups.  The remaining stars have photometric or
astrometric indications of their youth, and either reside on the
pre-main-sequence or upper main-sequence.

We then reduced TESS full frame images taken over the course of about
two months (Sectors 6 and 7; the first fields close to the galactic
plane).  We performed difference imaging to deal with the complex
background.  Using forced aperture photometry, we made light curves
for all Gaia-DR2 sources brighter than $G_{\rm Rp}$ of 13, and went
three magnitudes deeper for our target star sample.  This yielded
\numberlcs target star light curves across \numberclusters distinct
clusters.  The number of light curves per cluster was reported
(Tables~\ref{tbl:s6_lcs} and~\ref{tbl:s7_lcs}).  The software
developed for the reduction is available online
\citep{bhatti_cdips-pipeline_2019}.

The light curves seem to be limited in precision by photon-counting
noise from the target star at the bright end, and by unresolved
background stars at the faint end (Figure~\ref{fig:rms_vs_mag}).
Though the raw light curves show significant red noise, decorrelating
against a set of template stars led to an ensemble of light curves
with very nearly white noise properties
(Figures~\ref{fig:lc_systematics_dtr} and~\ref{fig:avg_acf}).

Brief exploration of the data revealed pulsating stars, eclipsing
binaries, and planet candidates (Figure~\ref{fig:quilt}).  A detailed
planet search is the subject of ongoing work.

We expect that our results will complement a number of other TESS data
processing efforts.  These include the NASA Ames SPOC pipeline
\citep{jenkins_spoc_2010}, the MIT Quick-Look-Pipeline
\citep{huang_tess_2018}, \texttt{eleanor}
\citep{feinstein_eleanor_2019}, the \citet{oelkers_precision_2018}
difference imaging pipeline, the TESS Asteroseismic Consortium (TASOC)
pipeline \citep{lund_k2p_2015,handberg_tess_2019}, and the Padova
team's effort (D{.} Nardiello, submitted; {\it e.g.},
\citealt{libralato_psf-based_2016}). 

Most of these pipelines are geared towards tasks more general than our
own, and most use aperture photometry.  The SPOC pipeline produces the
calibrated FFIs from raw images, processes 2 minute data, identifies
exoplanets, and produces the TESS data products of record.  The
MIT-QLP processes a subset of stars on the FFIs, identifies
exoplanets, and the affiliated TESS Science Office alerts object of
interest for ground and space-based followup.  The \texttt{eleanor}
pipeline is a user-friendly tool for extracting light curves from the
FFIs.  The team developing \texttt{eleanor} is also reporting planet
candidates to
ExoFOP-TESS\footnote{\url{exofop.ipac.caltech.edu/tess/index.php}}.
The TASOC pipeline is primarily aimed at asteroseismology and includes
classification modules for many types of variable stars. The
\citeauthor{oelkers_precision_2018} pipeline was aimed at broad,
all-sky variability searches, and used difference imaging methods
similar to our own.  Finally, the Padova team's effort is also
directed towards star clusters, but uses PSF subtraction to mitigate
crowding.  As light curves from these groups continue to be released,
this ecosystem should provide many opportunities to compare and
improve data analysis techniques. 

There are of course caveats and areas for improvement in our own work.
One methodological point concerns the hyperparameter tuning required
by our difference imaging method (\S~\ref{subsec:tuneconvkernel}).
Within our fine-tuning experiments over different kernel box-sizes and
polynomial weightings, we found that low-significance transits  can be
``lost'' for different choices of parameters that ideally should not
affect the photometric pipeline's results.  In the longer term,
developing an image-subtraction method that marginalizes over
uncertainties of how to chose ``optimal'' kernels would be desirable.
Pixel-level image subtraction methods that omit these parameters
entirely are also worth exploring \citep{wang_pixel-level_2017}.

A separate point to re-emphasize is that stars in our sample must be
understood to be {\it candidate} cluster stars, and ruling out the
possibility of photometric blending is important in subsequent vetting
efforts of any variable object.  Careful understanding of the cluster
itself is also important, as some clusters are less certain to exist
than others \citep[{\it e.g.}, the infrared clusters described
by][]{froebrich_FSR_2007}.

We remind the reader that our goal in creating our target catalog was
completeness, rather than reliability.  To create clean sub-samples,
we advise using the \texttt{CDEXTCAT} header keyword available in the
FITS files, which can be merged against the original source catalog to
obtain the membership probabilities reported by the original authors.
Alternatively, simply restricting the targets of interest to those
with provenance from {\it e.g.}, the Gaia-DR2 data is another way to
produce a clean sample of cluster targets.

Despite our goal of completeness, some clusters still may be missing
members~--~the census of nearby coeval stellar populations is very
much in flux. For instance, during the preparation of this work,
\citet{sim_open_2019} identified 209 new open cluster candidates
within 1~kpc through visual inspection of Gaia-DR2 data.  Similarly,
\citet{kounkel_untangling_2019} searched for groups near the galactic
plane within 1~kpc, and along with known clusters found hundreds of
new ``strings'' of kinematically associated stars, which could be
coeval.

Another area in which we may be incomplete is in sub-clusters of
large star-forming complexes. One important example is the Orion
Nebula (Messier~42; NGC~1976).  While it was observed by TESS in Sector~6,
searching our light curves for stars within 40 arcminutes and 100
parsecs of the Orion Nebula's center yielded only 180 light curves,
with 85 labelled members.  The Orion Nebula has far more known members
\citep{jones_proper_1988}.  However, due to a combination of the
spatially distributed nature of the broader complex, as well as
differential extinction, the automated methods of the
\citet{Kharchenko_et_al_2013} and \citet{dias_proper_2014} assigned
the Orion Nebula only 44 and 326 members, respectively.
\citet{cantat-gaudin_gaia_2018} explicitly excluded young star-forming
regions from their search, since the underlying assumption of their
clustering method (uniformity in the field star distribution) breaks
down in highly clustered star-forming regions.

In future work, these concerns will likely drive us to expand beyond
the current sample of target stars.  For the time being, the light
curves are of sufficient quality and quantity to begin astrophysical
studies.  We invite any who wish to explore the time evolution of
stellar or exoplanetary systems to interact with the
data\footnote{\stscilink} at \datasetlink.

\acknowledgements
L.G.B.\ acknowledges helpful discussions with 
T.~Cantat-Gaudin,
C.~Huang,
M.~Soares-Furtado,
J.~Wallace, and
S.~Yee.  The authors are also
grateful to the many people who have turned TESS from an idea into
reality.
L.G.B. and J.H. acknowledge support by the TESS GI Program, program
G011103, through NASA grant 80NSSC19K0386.
G.B. acknowledges support through NASA grant NNG14FC03C.
This paper includes data collected by the TESS mission, which are
publicly available from the Mikulski Archive for Space Telescopes
(MAST).
Funding for the TESS mission is provided by NASA's Science Mission
directorate.
This research has made use of the NASA Exoplanet Archive, which is
operated by the California Institute of Technology, under contract
with the National Aeronautics and Space Administration under the
Exoplanet Exploration Program.
This work made use of NASA's Astrophysics Data System Bibliographic
Services.
This research has made use of the VizieR catalogue access tool, CDS,
Strasbourg, France. The original description of the VizieR service was
published in A\&AS 143, 23.
This work has made use of data from the European Space Agency (ESA)
mission {\it Gaia} (\url{https://www.cosmos.esa.int/gaia}), processed
by the {\it Gaia} Data Processing and Analysis Consortium (DPAC,
\url{https://www.cosmos.esa.int/web/gaia/dpac/consortium}). Funding
for the DPAC has been provided by national institutions, in particular
the institutions participating in the {\it Gaia} Multilateral
Agreement.
The Digitized Sky Surveys were produced at the Space Telescope Science
Institute under U.S. Government grant NAG W-2166. The images of these
surveys are based on photographic data obtained using the Oschin
Schmidt Telescope on Palomar Mountain and the UK Schmidt Telescope.
This research has made use of the Exoplanet Follow-up Observation
Program website, which is operated by the California Institute of
Technology, under contract with the National Aeronautics and Space
Administration under the Exoplanet Exploration Program.
\newline
\facility{
	2MASS \citep{skrutskie_tmass_2006},
	Gaia \citep{gaia_collaboration_gaia_2016,gaia_collaboration_gaia_2018},
	TESS \citep{ricker_transiting_2015},
	UCAC4 \citep{zacharias_fourth_2013}
}
\software{
  \texttt{astrobase} \citep{bhatti_astrobase_2018},
  \texttt{astropy} \citep{the_astropy_collaboration_astropy_2018},
  \texttt{astroquery} \citep{astroquery_2018},
  \texttt{BATMAN} \citep{kreidberg_batman_2015},
  \texttt{cdips-pipeline} \citep{bhatti_cdips-pipeline_2019}
  \texttt{corner} \citep{corner_2016},
  \texttt{emcee} \citep{foreman-mackey_emcee_2013},
  \texttt{fitsh} \citep{Pal_2012},
  \texttt{IPython} \citep{perez_2007},
  \texttt{matplotlib} \citep{hunter_matplotlib_2007}, 
  \texttt{numpy} \citep{walt_numpy_2011}, 
  \texttt{pandas} \citep{mckinney-proc-scipy-2010},
  \texttt{pyGAM} \citep{serven_pygam_2018_1476122}
  \texttt{psycopg2} (\url{initd.org/psycopg})
  \texttt{scipy} \citep{jones_scipy_2001},
  \texttt{scikit-learn} \citep{sklearn_2011},
  \texttt{TagSpaces} (\url{tagspaces.org}),
  \texttt{tesscut} \citep{brasseur_astrocut_2019},
  \texttt{VARTOOLS} \citep{Hartman_Bakos_2016}
  \texttt{wotan} \citep{hippke_wotan_2019},
}

\begin{deluxetable*}{lllll}
	
	\tabletypesize{\footnotesize}
	
    \tablecaption{
    	CDIPS target star catalog assembled for this work. 
    	\label{tbl:cdips_targets}
	}
    \tablenum{1}
    
    \tablehead{
      \colhead{ID} &
      \colhead{497093746702988672} &
      \colhead{2006361919000756992} &
      \colhead{2048442943413035776} & 
      \colhead{5867618572762132864}
    }

    \startdata
     Cluster &                                                 Platais 3 &             ASCC 121 &           Teutsch 35 &           Loden\_1152 \\
 Reference &                                                Dias2014 &             Dias2014 &             Dias2014 &       Kharchenko2013 \\
 Ext catalog name &                                       799-012054 &           724-092321 &           630-074483 &            650278772 \\
 RA &                                                        71.4458 &              337.708 &              294.081 &              209.832 \\
 Dec &                                                       69.6454 &              54.6607 &              35.8782 &             -59.2883 \\
 PMra &                                                      -5.1826 &             -3.48356 &             -3.19716 &              -3.6399 \\
 PMdec &                                                   -0.478063 &             -1.28297 &             -2.04354 &             -3.93699 \\
 Plx &                                                      0.579745 &             0.196117 &             0.324896 &             0.397703 \\
 $G$ &                                                       15.1735 &              15.4843 &              15.1863 &              16.4277 \\
 $G_{\rm Bp}$ &                                              15.5786 &              16.3805 &              15.7505 &               16.991 \\
 $G_{\rm Rp}$ &                                              14.6008 &              14.5227 &              14.4853 &              15.6807 \\
 K13 name match &                                          Platais\_3 &             ASCC\_121 &             ASCC\_103 &           Loden\_1152 \\
 Unique cluster name &                                     Platais\_3 &             ASCC\_121 &             ASCC\_103 &           Loden\_1152 \\
 How match &                                            string\_match &         string\_match &         string\_match &         string\_match \\
 Not in K13? &                                                 False &                False &                False &                False \\
 Comment &                                                       NaN &                  NaN &                  NaN &                  NaN \\
 K13 logt &                                                      8.8 &                  6.4 &                 8.39 &                8.065 \\
 K13 err logt &                                                  NaN &                  NaN &                  NaN &                0.092 \\

    \enddata
    
    \tablecomments{
    This table is published in its entirety in a 
    machine-readable format.
    A portion of the transposed version
    is shown here for guidance regarding its form and content.
    Each row in the machine-readable version (each column in this
    version) represents a target star.
    The unique identifier, ``ID'', is the Gaia-DR2 source identifier.
    The external catalog(s) claiming cluster membership is given 
    as a comma-separated string in
    ``Reference'', and the name they assign is given as a comma-separated
    string in ``Cluster''.
    ``Ext catalog name'' is the name the external catalog assigns.
    Positions, proper motions, and the parallax are from Gaia-DR2.
    The magnitudes in Gaia $G$, $G_{\rm Bp}$, and $G_{\rm Rp}$ bands
    are given.
    The name matching described in Appendix~\ref{appendix:uniquenames}
    most often succeeds in finding the
    \citet{Kharchenko_et_al_2013} (K13) cluster corresponding to
    the external catalog claiming membership.
    This, or else the external name is used to assign the unique
    cluster name.
    The method for name matching (Appendix~\ref{appendix:uniquenames}) is
    also given as a string, as is a ``Comment'' summarizing
    information from \citet{Kharchenko_et_al_2013} about the
    cluster.
    The age and error as quoted by \citet{Kharchenko_et_al_2013}
    are also given.
}
 
\end{deluxetable*}

\begin{deluxetable}{lll}
	\tablecaption{Counts of light curves per cluster in Sector 6, sorted
		in descending order.  \label{tbl:s6_lcs}}
	\tablenum{2}
	
	\tablehead{
		\colhead{\hspace{.9cm}Name}\hspace{.9cm}
		& \colhead{\hspace{.9cm}$N_{\rm lc}$}\hspace{.9cm}
		& \colhead{\hspace{.9cm}Description}\hspace{.2cm}
	}
	
	\startdata
	     Platais\_5 &               7074 &      =,m,o, \\
     Platais\_6 &               7016 &      =,m,o, \\
     Mamajek\_3 &               3498 &      =,m,o, \\
  Collinder\_70 &               1576 &      =,a,o, \\
    Trumpler\_5 &               1492 &     =,,,var \\
  Collinder\_69 &               1192 &        =,,, \\
 Collinder\_110 &               1110 &        =,,, \\
       ASCC\_21 &               1094 &      =,a,c, \\
 Collinder\_121 &                986 &      =,a,o, \\
       ASCC\_19 &                925 &        =,,, \\
      NGC\_2287 &                869 &        =,,, \\
      NGC\_2232 &                760 &        =,,, \\
       ASCC\_20 &                756 &        =,,, \\
      NGC\_2141 &                730 &        =,,, \\
      NGC\_2112 &                673 &        =,,, \\
       ASCC\_16 &                661 &     =,,,ass \\
      NGC\_2194 &                656 &        =,,, \\
      NGC\_2301 &                586 &        =,,, \\
  Collinder\_65 &                575 &        =,,, \\
       ASCC\_28 &                533 &        =,,, \\

	\enddata
	
	\tablecomments{
		Table~\ref{tbl:s6_lcs} is published in its entirety in a machine-readable format.
		The top twenty entries are shown here for guidance regarding form and content.
		Names are matched against \citet{Kharchenko_et_al_2013} as
		described in Appendix~\ref{appendix:uniquenames}, and $N_{\rm lc}$
		is the number of light curves associated with the cluster from this
		data release.
		The description column matches \citet{Kharchenko_et_al_2013},
		and is in the format ``\texttt{a},\texttt{b},\texttt{c},\texttt{d}''.
		Meanings of displayed symbols are as follows.
		\texttt{a}:
		``='' = cluster parameters were determined,
		``\&'' = duplicated/coincides with other cluster;
		\texttt{b}:
		``blank'' = open cluster,
		``a'' =	association,
		``m'' =	moving group,
		``n'' =	nebulosity;
		\texttt{c}:
		``o'' = object,
		``c'' = candidate;
		\texttt{d}:
		``ass'' =	stellar association,
		``var'' =	clusters with variable extinction.
	}
	
	\vspace{-1cm}
\end{deluxetable}

\begin{deluxetable}{lll}
	\tablecaption{Counts of light curves per cluster in Sector 7, sorted
		in descending order.  See Table~\ref{tbl:s6_lcs} for notes. \label{tbl:s7_lcs}}
	\tablenum{3}
	
	\tablehead{
		\colhead{\hspace{.9cm}Name}\hspace{.9cm}
		& \colhead{\hspace{.9cm}$N_{\rm lc}$}\hspace{.9cm}
		& \colhead{\hspace{.9cm}Description}\hspace{.2cm}
	}
	
	\startdata
	 Collinder\_173 &              13009 &   \&,a,o,ass \\
       ASCC\_33 &               2826 &      =,n,o, \\
      NGC\_2437 &               2365 &        =,,, \\
      NGC\_2477 &               2158 &        =,,, \\
      NGC\_2546 &               1616 &        =,,, \\
     NGC\_2451A &               1486 &        =,,, \\
     NGC\_2451B &               1288 &        =,,, \\
      NGC\_2516 &               1278 &        =,,, \\
      NGC\_2323 &               1239 &        =,,, \\
      NGC\_2447 &               1192 &        =,,, \\
 Collinder\_132 &               1074 &        =,,, \\
       ASCC\_32 &                995 &        =,,, \\
 Collinder\_121 &                976 &      =,a,o, \\
      NGC\_2360 &                894 &        =,,, \\
      NGC\_2287 &                879 &        =,,, \\
      NGC\_2506 &                877 &        =,,, \\
      NGC\_2548 &                807 &        =,,, \\
      NGC\_2539 &                726 &        =,,, \\
     Alessi\_21 &                704 &        =,,, \\
    Melotte\_71 &                630 &        =,,, \\

	\enddata
	
	\tablecomments{
		Table~\ref{tbl:s7_lcs} is published in its entirety in a machine-readable format.
		A portion is shown here for guidance regarding its form and content.
		Format is same as Table~\ref{tbl:s6_lcs}.
	}
	
	\vspace{-1cm}
\end{deluxetable}

\clearpage
\newpage

\bibliographystyle{yahapj}                            
\bibliography{bibliography} 

\appendix
\section{Time system \& barycentric correction}
\label{appendix:time}

The time-stamps included with the calibrated TESS Full Frame Images
produced by SPOC include a barycenteric correction at
a single reference pixel given at the middle of every frame.
The barycentric correction is at maximum 16 minutes, corresponding to
points on the sky separated by 180 degrees.
The angular distance from a TESS camera's center of field to the corners
is $\approx$17 degrees, so naively one might incur at worst an error of
$\approx$90 seconds on the time-stamps due to using a barycentric
correction in a direction that is slightly wrong.
Nonetheless, following
\citet{bouma_wasp-4b_2019},
we perform our own barycentric correction using the appropriate
sky coordinates for each light curve.
We advise use of our \texttt{TMID\_BJD} column, which gives the
mid-time of each exposure in the BJD$_{\rm TDB}$ time system, which
is the defacto standard in exoplanet and stellar
astronomy~\citep{eastman_achieving_2010}.

\section{Assigning unique names to each cluster}
\label{appendix:uniquenames}

In assigning a single unique cluster name to each star, we matched
against the \citet{Kharchenko_et_al_2013} name whenever possible,
since this was the largest available catalog, and it also included
homogeneous age determinations for many of the clusters.
To find the matching name, in order of precedence we
\begin{enumerate}
  \item Checked for direct string matches from
    \citet{Kharchenko_et_al_2013} clusters with determined parameters;
  \item Checked whether the SIMBAD online name resolving service \citep{wenger_simbad_2000} had
    any direct string matches against \citet{Kharchenko_et_al_2013}
    clusters with determined parameters;
  \item Checked for string matches in the full
    \citet{Kharchenko_et_al_2013} index (including clusters without
    determined parameters);
  \item Searched for spatial matches between each star and cluster
    centers from \citet{Kharchenko_et_al_2013} within 10 arcminutes.
    In cases with multiple cluster matches, we ignored candidate
    matches to avoid assigning incorrect names;
  \item Checked the WEBDA double name
    list\footnote{\url{https://webda.physics.muni.cz/double_names.html}, accessed \texttt{2019-08-12}},
    and repeated Steps 1-4 with any matches.
\end{enumerate}

A few edge-cases, including sub-clusters of larger
star-forming complexes like in Sco-Cen or Collinder~33, were 
manually resolved to the extent feasible (\citealt{rizzuto_multidimensional_2011}
and \citealt{saurin_isolating_2015} give
detailed pictures of the complex morphologies that frequently arise in
young star-forming regions).

The procedure described above failed to yield matches for a few of the
infrared clusters identified by \citet{majaess_discovering_2013} and
included in the \citet{dias_proper_2014} catalog.
For these cases, we used the name given by \citet{dias_proper_2014}.
The larger set of ``FSR'' infrared clusters from
\citet{froebrich_FSR_2007} was incorporated to
\citet{Kharchenko_et_al_2013}, and so did not present any
complications.

The Hyades and a number of other nearby moving groups were also
missed, since they were not in the \citet{Kharchenko_et_al_2013}
catalog.  For moving groups not identified in
\citet{Kharchenko_et_al_2013}, we adopted the constellation-based
naming convention from \citet{gagne_banyan_XI_2018}.  

Finally, the procedure enumerated above did not yield matches for
recently discovered clusters, such as the ``RSG'' clusters found by
\citet{roser_nine_RSG_2016} and the ``Gulliver'' clusters from
\citet{cantat-gaudin_gaia_2018}.  In these cases, we used the names
given by the original authors.

\listofchanges
\end{document}